\documentclass[a4paper,UKenglish,numberwithinsect]{lipics-v2021}
\pdfoutput=1

\usepackage{cite} 
\usepackage{amsmath}
\usepackage{amssymb}  
\usepackage{xspace}
\usepackage{enumerate}
\usepackage{subcaption}
\usepackage{complexity}
\usepackage{caption}
\usepackage{amsthm}
\usepackage[numbers]{natbib}
\usepackage[capitalize]{cleveref}

\DeclareMathOperator{\crn}{cr}

\usepackage{tikz}
\usepackage{todonotes}
\tikzstyle{every picture} = [>=latex]
\usetikzlibrary{math, calc, arrows, quotes}
\usetikzlibrary {decorations.pathmorphing, decorations.pathreplacing, decorations.shapes, decorations.markings}
\usetikzlibrary{backgrounds}

\newcommand{\HB}{\texttt{HB}}
\newcommand{\LB}{\texttt{LB}}
\newcommand{\Rr}{\texttt{R}}
\newcommand{\Rrh}{\texttt{R\hskip-1pt'}}
\newcommand{\Bb}{\texttt{B}}
\newcommand{\Bbh}{\texttt{B\hskip-1pt'}}
\newcommand{\Gg}{\texttt{G}}
\newcommand{\Cc}{\texttt{C}}
\newcommand{\true}{\texttt{True}}
\newcommand{\false}{\texttt{False}}
\newcommand{\calG}{\mathcal{G}}

\newtheorem{obs}[theorem]{Observation}

\DeclareMathOperator{\Var}{{\sf Var}}

\def\tw{\mathop{\mbox{\rm tw}}}
\def\crg{\operatorname{cr}} 
\def\ca#1{\mathcal{#1}}

\newcommand{\defquestion}[3]{
	\vspace{2mm}
	\noindent\fbox{
		\begin{minipage}{0.96\linewidth}
			\begin{tabular*}{\linewidth}{@{\extracolsep{\fill}}lr} \textsc{#1} & \\ \end{tabular*}
			{\bf{Input:}} #2 \\
			{\bf{Question:}} #3
		\end{minipage}
	}
	\vspace{2mm}
}

 \usepackage[appendix=inline]{apxproof}
\ifx\proof\inlineproof

\else

\fi

\theoremstyle{claimstyle}

\def\tw{\mathop{\mbox{\rm tw}}}
\def\crg{\operatorname{cr}} 
\def\ca#1{\mathcal{#1}}

\newif\ifblindre
\ifblindre\else
\hideLIPIcs
\nolinenumbers
\fi

\tikzstyle{uweight}=[very thin, decorate,decoration={zigzag,segment length=4,amplitude=0.4}]
\def\blackpart{!75!black}
\def\gblackpart{!65!black}  
\tikzstyle{framebars}=[brown!50!black,ultra thick]
\tikzstyle{framebars2}=[brown!50!black,ultra thick]
\tikzstyle{varibars}=[black,thick]

\def\tikzonevar#1#2#3{%
	\tikzmath{integer \y; \y = #2+1;}
	\tikzstyle{every node}=[draw, color=red\blackpart, shape=circle, inner sep=0.8pt, fill=red\blackpart]
	\tikzstyle{every path}=[draw, color=red\blackpart]
	\draw (#1-.5,#2+3) node (l\y#3) {} (#1+.5,#2+3) node (r\y#3) {} (#1-.5,1) node (l-1#3) {} (#1+.5,1) node (r-1#3){};
	\draw[thick] (r-1#3) --++(1,1) node (r0#3){} --++(0,#2) node (r#2#3){} -- (r\y#3);
	\draw[uweight,semithick] (r-1#3) --(r0#3) -- (r#2#3) -- (r\y#3);
	\draw[thick] (l-1#3) --++(-1,1) node (l0#3){} --++(0,#2) node (l#2#3){} -- (l\y#3);
	\draw[uweight,semithick] (l-1#3) --(l0#3) -- (l#2#3) -- (l\y#3);
	\foreach \xx in {2,...,#2} { \tikzmath{integer \x; \x = \xx-1;}
		\node (r\x#3) at (#1+1.5,\x+2) {};  \node (l\x#3) at (#1-1.5,\x+2) {};
	}
	
	\tikzstyle{every node}=[draw, color=blue\blackpart, shape=circle, inner sep=0.8pt, fill=blue\blackpart]
	\tikzstyle{every path}=[draw, color=blue\blackpart]
	\draw (#1-2,#2+3) node (ll\y#3) {} (#1+2,#2+3) node (rr\y#3) {} (#1-2,1) node (ll0#3) {} (#1+2,1) node (rr0#3) {};
	\draw[thick] (rr0#3) -- (rr\y#3) (ll0#3) -- (ll\y#3);
	\draw[uweight,semithick] (rr0#3) -- (rr\y#3) (ll0#3) -- (ll\y#3);
	\foreach \x in {1,...,#2} {
		\node (rr\x#3) at (#1+2,\x+1.5) {};  \node (ll\x#3) at (#1-2,\x+1.5) {};
		\draw (ll\x#3)--++(1,0) (rr\x#3)--++(-1,0);
	}
	
	\tikzstyle{every node}=[draw, varibars, shape=circle, inner sep=1pt, fill]
	\tikzstyle{every path}=[draw, framebars]
	\draw (ll0#3) node{} -- (#1,0) node (ad#3) {} -- (rr0#3) node{} ;
	\draw (#1-1,2) node (bd#3) {} -- (#1+1,2) node (cd#3) {};
	\draw (ad#3) -- (r-1#3) node{} --++(-0.5,1) node (add#3) {} -- (l-1#3) node{} -- (ad#3);
	\draw (ll\y#3) node{} -- (#1,#2+4) node (au#3) {} -- (rr\y#3) node{} ;
	\draw (#1-1,#2+2) node (bu#3) {} -- (#1+1,#2+2) node (cu#3) {};
	\draw (au#3) -- (r\y#3) node{} --++(-0.5,-1) node (auu#3) {} -- (l\y#3) node{} -- (au#3);
	\tikzstyle{every path}=[draw, varibars]
	\draw (bd#3)--(bu#3) (cd#3)--(cu#3);
	\tikzstyle{every node}=[draw, color=black, shape=circle, inner sep=0.8pt, fill=black]
	\foreach \x in {1,...,#2} {
		\node (c\x#3) at (#1+1,\x+1.5) {};  \node (b\x#3) at (#1-1,\x+1.5) {};
	}
	
	\tikzstyle{every node}=[draw, color=blue\blackpart, shape=circle, inner sep=1pt, fill=blue\blackpart]
	\draw (#1-2,#2+3) node (ll\y#3) {} (#1+2,#2+3) node (rr\y#3) {} (#1-2,1) node (ll0#3) {} (#1+2,1) node (rr0#3) {};
	
	\tikzstyle{every node}=[draw, color=red\blackpart, shape=circle, inner sep=1pt, fill=red\blackpart]
	\draw (#1-.5,#2+3) node (l\y#3) {} (#1+.5,#2+3) node (r\y#3) {} (#1-.5,1) node (l-1#3) {} (#1+.5,1) node (r-1#3){};
}

\def\slightup{--++(1.25,0) --++(1.5,0.5) -- ++(0.75,0) --++(0.5,0.5) --++(1,0)}
\def\jumponeup{--++(0.5,0) --++(1.5,1.5) -- ++(1.5,0) --++(0.5,0.5) --++(1,0)}
\def\slightupend{--++(0.5,0.5) --++(2.5,0) --++(1,0.5) --++(1.75,0)}
\def\jumponeupend{--++(0.5,0) --++(1.5,1.5) -- ++(1.5,0) --++(0.75,0)}

\title{Crossing Number is \NP-hard for Constant Path-width (and Tree-width)\thanks{An extended abstract of a preliminary version of this paper was presented at ISAAC'24~\cite{HK-isaac24}.}}
\titlerunning{Crossing Number is \NP-hard for Constant Path-width (and Tree-width)}

\ifblindre
\author{Anonymous I}{}{}{}{}
\author{Anonymous II \dots}{}{}{}{}
\authorrunning{Anonymous \dots}
\else
\author{Petr Hlin{\v e}n\'y}{Masaryk University, Brno, Czech Republic \and \url{https://www.fi.muni.cz/~hlineny}}{hlineny@fi.muni.cz}{https://orcid.org/0000-0003-2125-1514}{Czech Science Foundation [project 26-21334S].}
\author{Liana Khazaliya}{Technische Universit\"{a}t Wien, Vienna, Austria \and \url{https://www.ac.tuwien.ac.at/people/lkhazaliya/}}{lkhazaliya@ac.tuwien.ac.at}{https://orcid.org/0009-0002-3012-7240}{Vienna Science and Technology Fund (WWTF) [10.47379/ICT22029]; Austrian Science Fund (FWF) [10.55776/Y1329]; European Union's Horizon 2020 COFUND programme [LogiCS@TUWien, grant agreement No.\ 101034440].}
  
\authorrunning{P.\ Hlin\v{e}n\'y and L.~Khazaliya}
\Copyright{Petr Hlin\v{e}n\'y and Liana Khazaliya}
\fi



\begin{document}
\maketitle

\begin{abstract}
The \emph{crossing number} of a graph is the minimum number of edge crossings that a graph can have when drawn in the plane. 
Determining this number, known as the~\textsc{Crossing Number} problem, is a celebrated problem in combinatorial optimization.
It has been known to be \NP-complete since the~1980s, and already showing its fixed-parameter tractability when parameterized by the vertex cover number required fairly involved techniques.
In this paper, we prove that computing the crossing number exactly remains \NP-hard even for graphs of path-width~$12$ (and as a result, for simple graphs of path-width~$13$ and tree-width~$9$).

These results highlight that, although both path- and tree-decompositions have been highly successful tools in many graph algorithm scenarios,
general crossing number computation is unlikely (under~$\P\neq \NP$) to be successfully tackled using graph decompositions of bounded width---a question that had remained a~{'tantalizing open problem'} [S.~Cabello, Hardness of Approximation for Crossing Number, 2013] till now.
\end{abstract}

\section{Introduction}
The notion of the \emph{crossing number} originally arose during WWII in work of Tur\'an~\cite{Turan77}, who studied drawings of complete bipartite graphs, motivated by the problem of minimizing the number of crossings among tracks connecting brick kilns to storage sites. 
The \emph{crossing number}~$\crn(G)$ of a graph~$G$ is the minimum possible number of pairwise edge crossings over all drawings of $G$ in the plane; see \cref{sec:prelim}.
Determining the crossing number of a graph, known as the~\textsc{Crossing Number} problem, is one of the most prominent combinatorial optimization problems in graph theory.

In the 1980s, Garey and Johnson~\cite{GareyJ83} showed that the crossing number minimization problem is \NP-hard.
This hardness result has since been strengthened to hold even for fairly restrictive graph classes.
In particular, the problem remains \NP-hard for cubic graphs~\cite{DBLP:journals/jct/Hlineny06a}, as well as when a rotation scheme is fixed~\cite{DBLP:conf/gd/PelsmajerSS07a}.
Moreover, the~\textsc{Crossing Number} problem is \APX-hard~\cite{Cabello13} (does not admit a polynomial-time approximation scheme unless~$\P = \NP$) in its general setting.

Another major direction of extensive research concerns the computation of the crossing number for graphs that are initially close to being planar.
Surprisingly, the~\textsc{Crossing Number} remains \NP-hard for \emph{almost planar} graphs, i.e.,~graphs of the form~$(G + e)$ obtained by adding a single edge $e$ to a planar graph~$G$~\cite{DBLP:journals/siamcomp/CabelloM13}.
Consequently, with respect to each of the maximum degree of the graph and the number of edges whose removal makes the graph planar, the \textsc{Crossing Number} problem is para-\NP-hard.
Furthermore, in contrast to the remarkable polynomial-time solvable case of the \textsc{Crossing Number} for a graph~$(G + e)$ where~$G$ is planar of maximum degree three~\cite{riskin1996crossing,DBLP:journals/algorithmica/CabelloM11}, 
the problem remains \NP-hard for almost planar graphs $(G + e)$ when only three of the vertices of~$G$ have degree greater than~$3$~\cite{DBLP:conf/mfcs/Hlineny25}.

From the viewpoint of approximations, computing a~$(1+\epsilon)$-approximation of the crossing number is \NP-hard for some fixed~$\epsilon > 0$ by aforementioned~\cite{Cabello13}.
Although no non-trivial approximation results are known for general graphs, algorithms with bounded approximation ratios exist for graphs of bounded degree~\cite{DBLP:conf/focs/ChuzhoyMT20, DBLP:conf/stoc/ChuzhoyT22}, and more specifically for graphs embeddable on a fixed surface~\cite{DBLP:journals/jct/ChimaniHS20, DBLP:journals/combinatorics/GitlerHLS08} and almost-planar graphs~\cite{DBLP:journals/algorithmica/CabelloM11, DBLP:journals/jco/ChimaniH17, DBLP:conf/gd/HlinenyS06}.

A notable exception to the general intractability of an exact solution to the~\textsc{Crossing Number} problem is that it can be efficiently, in FPT, computed when the solution value is bounded. 
More precisely, one can decide whether there is a solution with at most $k$ crossings in time~$f(k)\cdot n^{\ca O(1)}$, in case of the ordinary \textsc{Crossing Number} and also for many other variants of the problem~\cite{DBLP:journals/jcss/Grohe04, DBLP:conf/stoc/KawarabayashiR07, cm-faeg2-21, DBLP:conf/compgeom/HammH22, DBLP:conf/esa/VerdiereH25}. 
Very recently, in line with this research, Lokshtanov et al.~\cite{DBLP:conf/soda/LokshtanovP0S0Z25} proposed an algorithm with running time~$2^{\mathcal{O}(k \log k)} \cdot n$ for the ordinary \textsc{Crossing Number}. 
On the negative side, the problem unlikely (unless~$\coNP \subseteq \NP/\poly$) admits a polynomial kernel~\cite{DBLP:conf/compgeom/HlinenyD16}.

Yet another traditional approach to addressing the computational hardness of graph problems is to exploit structural properties of the input graph in order to understand how these properties affect the computational feasibility of the problem.
From this perspective, it is known that if the input graph has a vertex cover of bounded size, then its crossing number can be computed exactly in fixed-parameter tractable time, i.e., in time~$f(k)\cdot n^{\mathcal{O}(1)}$, where~$k$ is the vertex cover number~\cite{DBLP:conf/gd/HlinenyS19}.
Accordingly, the study of the~\textsc{Crossing Number} with respect to other structural parameters---most notably feedback vertex set number, tree-depth, path-width, and tree-width---has been repeatedly mentioned as an interesting research venue to explore~\cite{DBLP:journals/algorithmica/BiedlCDM20, Cabello13, DBLP:journals/csr/Zehavi22}.
In this direction, it is known that the problem is solvable in linear time on maximal graphs of path-width~$3$, admits a~$2$-approximation algorithm on general graphs of path-width~$3$, and admits a~$4w^3$-approximation on maximal graphs of path-width~$w$~\cite{DBLP:journals/algorithmica/BiedlCDM20}.
For a comprehensive overview of tractability results on the~\textsc{Crossing Number} problem, we refer the reader to the recent survey by Zehavi~\cite{DBLP:journals/csr/Zehavi22}.

In this paper, we present a new hardness result: the~\textsc{Crossing Number} problem is \NP-hard even on graphs of constant path-width (and, respectively, tree-width), namely, for path-width~$12$ (and tree-width~$9$). 
This result immediately implies that the problem is para-\NP-hard, and resolves (assuming~$\P \not= \NP$) the question of whether the~\textsc{Crossing Number} problem possibly is in~\FPT\ or~\XP\ on the graph classes of bounded tree-width or path-width.

\begin{theorem}[cf.~\Cref{thm:mainred} and \Cref{cor:mainred}]\label{thm:main}
	Given a graph~$G$ and an integer~$k$, the problem of deciding whether~$G$ can be drawn in the plane with at most~$k$ crossings is \NP-complete, even when~$G$ is required to have path-width at most~$12$, or when~$G$ is required to be simple of path-width at most~$13$ or tree-width at most~$9$.
\end{theorem}

In~\cref{sec:prelim} we introduce the basic concepts and necessary preliminaries, including the definitions of a drawing, the crossing number, width decompositions, and the~\textsc{Crossing Number} problem itself.
In~\cref{sec:reduction} (\cref{thm:mainred}), we present a hardness reduction from \textsc{Satisfiability}.
Due to the technical nature of the proof, we separate it into several parts: the construction of the graph~(\cref{sub:fullred}), necessary conditions for the existence of a drawings with a predefined number of crossings~(\cref{subsec:claims}), correctness of the reduction~(\cref{subsec:correctness}), and finally, the verification that the width parameters---path-width and tree-width---of the constructed graphs remain constant~(\cref{subsec:width}).
We conclude with a discussion in~\cref{sec:conclu}.

\section{Preliminaries}\label{sec:prelim}
Throughout the paper, we consider finite graphs that may contain parallel edges.
We begin with standard terminology of graph theory~\cite{Diestel12}, including the notions of tree-width and path-width~\cite{CyganFKLMPPS15}, which are commonly used parameters to capture the structural complexity of a graph, as well as graph drawing concepts~\cite{BattistaETT99}.

For integers $n,k \in \mathbb{N}$, we write $[n] = \{1, \dots, n\}$ and $[k,n] = \{k, \dots, n\}$.

\subsection{Drawings}\label{subsec:drawings}

A \emph{drawing}~$\mathcal{G}$ of a graph~$G$ in the plane is a mapping of the vertices~$V(G)$ to distinct points in the plane, and of the edges~$E(G)$ to simple curves connecting their respective endpoints and containing no other vertex point.
When convenient, we will refer to the elements~(vertices and edges) of the drawing by the corresponding elements of~$G$.
A~\emph{crossing} is an intersection point of two distinct edge curves other than a common endpoint.
It is well established that, when searching for an optimal solution to the \textsc{Crossing Number} problem, one may restrict attention to so-called \emph{good drawings}: any two edges cross at most once, adjacent edges do not cross, and no three edges cross in a common point.

A drawing~$\mathcal{G}$ is \emph{planar} (or a \emph{plane graph}) if~$\mathcal{G}$ has no crossings, and a graph is \emph{planar} if it admits a planar drawing.
The number of crossings in a particular drawing~$\mathcal{G}$ is denoted by~$\crn(\mathcal{G})$, and the minimum of~$\crn(\mathcal{G})$ taken over all good drawings~$\mathcal{G}$ of a graph~$G$ is denoted by~$\crn(G)$.
We call~$\crn(\mathcal{G})$ and~$\crn(G)$ the \emph{crossing number} of the drawing~$\mathcal{G}$ and the graph~$G$, respectively.
The \textsc{Crossing Number} problem asks, for a given graph~$G$, to find a good drawing $\mathcal{G}$ with the minimum possible number of crossings.

We also use a standard artifice in crossing number research.
In a \emph{weighted} graph, each edge is assigned a positive number, called its \emph{weight} (or \emph{thickness}), typically an integer.
The \emph{weighted crossing number} is defined analogously to the ordinary crossing number, except that a crossing between two edges~$e_1$ and~$e_2$ of weights~$t_1$ and~$t_2$, respectively, contributes~$t_1 \cdot t_2$ to the total.

For the purpose of computing the crossing number, an edge of integer weight~$t$ can equivalently be replaced by~$t$ parallel edges of weights~$1$, since these~$t$ edges can be drawn arbitrarily close to one another without increasing the number of crossings.
Consequently, we henceforth use weighted edges instead of parallel edges and, unless stated otherwise, refer simply to the \emph{crossing number} to mean the weighted crossing number.

\subsection{Tree-width and Path-width}\label{sub:twpw}

A \emph{tree decomposition}~$\mathcal{T}$ of an undirected graph~$G$ is a pair~$(T,\chi)$, where~$T$ is a tree whose vertices we called \emph{nodes} and~$\chi$ is a function that assigns to each node~$t \in V(T)$ a set~$\chi(t) \subseteq V(G)$, such that the following conditions holds: 
\begin{itemize}
	\item For every edge~$\{u, v\} \in E(G)$, there exists a node~$t\in V(T)$ such that~$u,v\in \chi(t)$.
	\item For every vertex~$v \in V(G)$, the set of nodes~$t\in V(T)$ with~$v\in \chi(t)$ induces a nonempty subtree of~$T$.
\end{itemize}
The sets~$\chi(t)$, for~$t \in V(T)$, are called the \emph{bags} of the tree decomposition.
The \emph{width} of a tree decomposition~$(T,\chi)$ is~$\max_{t\in V(T)}|\chi(t)|-1$, and the \emph{tree-width} of the graph~$G$, denoted~$\tw(G)$, is the minimum width over all tree decompositions of~$G$.

The \emph{path decomposition} and the corresponding notion of \emph{path-width} are defined analogously, with the only difference that the tree~$T$ is required to be a path.

We consider the characterization of a graph~$G$ using the {\em cops-and-robber game}.
The game is played between the {\em cops} and the {\em robber}, and the rules are as follows:
\begin{itemize}
	\item The {\em robber} may move freely along paths of~$G$ that contain no cop.
	\item The {\em cops} are transported by helicopter and may land on any vertex or be lifted back up.
	When a helicopter appears above a vertex with the robber, the robber may move along any path that contains no cop before the cop lands.
	\item The robber is caught whenever a cop lands on the robber's vertex currently occupied by the robber.
\end{itemize}
Such a game is called \emph{monotone} if the robber is never able to reach a vertex that was previously occupied by a cop.

The cited characterization is the following.
\begin{theorem}[Seymour and Thomas \cite{DBLP:journals/jct/SeymourT93}]\label{thm:STcops}~
	\begin{enumerate}[(1)]
		\item The tree-width of~$G$ is at most~$t$ if and only if~$t+1$ cops can always catch the robber in a monotone game on~$G$ when the robber is visible to the cops.
		\item The path-width of~$G$ is at most~$t$ if and only if~$t+1$ cops can always catch the robber in a monotone game on~$G$ when the robber is \emph{not} visible to the cops.
	\end{enumerate}
\end{theorem}

\section{Hardness Reduction}\label{sec:reduction}

In this section, we describe and prove a polynomial-time reduction that, given an instance~$\mathcal{I}=(\mathcal{C}, \mathcal{V})$ of \textsc{Satisfiability}, outputs an equivalent instance~$(G,k)$ of \textsc{Crossing Number} on a graph of constant path-width (and tree-width). 

\defquestion
{Satisfiability}
{A set of clauses~$\mathcal{C}=\{C_1, \dots, C_{\ell}\}$ over variables~$\mathcal{V}= \{x_1, \dots, x_n\}$}
{Does there exist an assignment of the variables~$\tau: \mathcal{V} \rightarrow \{\true, \false\}$ that satisfies all clauses in~$\mathcal{C}$?}

\defquestion{Crossing Number}{A graph~$G$ and an integer~$k\in \mathbb{Z}_{\geq0}$}{Does $G$ admit a drawing~$\mathcal{G}$ in the plane with at most~$k$ crossings?}

\begin{theorem}\label{thm:mainred}
	There exists a polynomial-time algorithm that, given an instance~$\mathcal{I}$ of \textsc{Satisfiability}, 
	outputs an equivalent instance $(G,k)$ of \textsc{Crossing Number} such that the graph~$G$ is of path-width at most~$12$ and tree-width at most~$9$.
\end{theorem}
We remark that the graph $G$ in \cref{thm:mainred} may (and will) contain parallel edges.
If simplicity of the graph~$G$ is desired, we immediately obtain the following corollary:
\begin{corollary}\label{cor:mainred}
	There exists a polynomial-time algorithm that, given an instance~$\mathcal{I}$ of \textsc{Satisfiability}, 
	outputs an equivalent instance $(G,k)$ of \textsc{Crossing Number} such that the graph~$G$ is simple of path-width at most~$13$ and tree-width at most~$9$.
\end{corollary}
\begin{proof}
	For any graph~$G$ and edge~$e\in E(G)$, the same drawing as a point set can be used for~$G$ and the graph obtained by subdividing~$e$.
	Obviously, subdivisions of edges do not change the crossing number of a graph.
	Hence, if the graph~$G$ of~\Cref{thm:mainred} contains parallel edges, we can form a graph~$G'$ by subdividing each such edge of~$G$ once; then~$\crg(G')=\crg(G)$.
	The tree-width remains unchanged, and the path-width of~$G'$ increases by at most~$1$ compared to~$G$.
\end{proof}

\subsection{High-level Idea}

Naturally, in interpreting an instance~$\mathcal{I}=(\mathcal{V}, \mathcal{C})$ of  \textsc{Satisfiability} as an instance~$(G, k)$ of \textsc{Crossing Number}, one might use a large grid-like structure.
Such a structure would allow the values of the variables in~$\mathcal{V}$ to be represented independently, while enabling each clause in~$\mathcal{C}$ to interact with its corresponding variables.
For example, one may imagine arranging variables as columns and clauses as rows of the grid, with their interactions encoded by gadgets located in specially crafted cell at the corresponding row-column intersections.

However, if a graph contains a large grid as a minor, then its tree-width is also large, whereas our goal is to obtain a graph~$G$ of constant tree-width and path-width.
We therefore base our reduction on a \emph{frame graph}~$F$ that admits many small separators (here of size~$4+4$) arranged in a left-to-right order, which allows us to ensure that the resulting graph~$G$ has constant path-width.
The crucial property of this construction is that, for each such separator~$X$, the graph~$(F-X)$ has three connected components---the \emph{left}, \emph{middle}, and \emph{right} components---such that the left and right components are forced to cross the middle component many times in any drawing whose total number of crossings is bounded by a predefined constant
(see \cref{fig:auxill} and \cref{fig:example-noflip} for an illustration).
In this way, we enforce the desired large grid-like structure in any optimal drawing of the frame~$F$, and consequently in any optimal drawing of~$G$.

At the same time, the frame is constructed so as to allow a certain limited form of drawing flexibility, namely the possibility of performing \emph{vertical flips} of the middle components associated with the separators described above (see~\cref{fig:auxill}).
These flips form the basis of the variable gadgets in our reduction.
We exploit this drawing flexibility to encode truth assignments of variables in \textsc{Satisfiability} (see \cref{fig:example-assignments} and \cref{fig:example-assignments-b} for an illustration of the encoding).
To encode the participation of variables in clauses, we attach specific small gadgets (see~\cref{fig:cells}) to the variable gadgets in~$G$.
A satisfying assignment is then verified by the existence of a drawing of the global clause edges of~$G$---one edge per clause, shown in green in~\cref{fig:example-noflip}---that incurs the minimum possible total crossing cost.
This approach is reminiscent of the construction used in~\cite{DBLP:journals/siamcomp/CabelloM13}.

The crucial aspect of our construction, however, lies in enforcing the unique correct crossing pattern between the components of the frame, as depicted in~\cref{fig:auxill}.
To achieve this, we build upon an idea originally introduced in~\cite{DBLP:conf/isaac/HlinenyS15}, which we develop and formalize in~\cref{subsec:claims}.

\subsection{Auxiliary Graphs}\label{sub:auxill}

To facilitate the presentation, we use colors---heavy-brown~(\HB), light-black~(\LB), red~(\Rr), blue~(\Bb), cyan~(\Cc), green~(\Gg)---to encode the future order of the weights of the corresponding edges (see their definition in~\cref{tab:colors}). 
The weights play a crucial role in the subsequent description of possible drawings of the constructed graph.
The weight values are defined with respect to a sufficiently large base weight~$\omega$, which is still polynomial in~$|\mathcal{C}|+|\mathcal{V}|$ (e.g.,~$\omega = |E(G)|^2$).
Informally, a single crossing of weight~$\omega^{t+1}$ ``outweighs'' all crossings of weight~$\omega^t$ in~$G$, ensuring that crossings of larger weight determine the structure of any optimal drawing.
\begin{table}[t]\renewcommand{\arraystretch}{1.2}\centering
	\begin{tabular}{||l l l||} 
		\hline
		Color& Usage & Weight  \\ 
		\hline\hline
		Heavy-brown~(\HB) & The frame and $\Var$-gadgets attachments & $\omega^{8}$ \\ 
		\hline
		Light-black~(\LB)& $\Var$-gadgets interior skeletons  & $\omega^{6}$ \\ 
		\hline
		Red~(\Rr) & Paths in $\Var$-gadgets (vertical) & $\omega^{4}+\Theta_{n,\ell}(\omega^{1})\,>\omega^{4}$\\ 
		\qquad(\Rrh)	 & Stairs interconnecting $\Var$-gadgets (horizontal) & $\omega^{3}$ \\ 
		\hline
		Blue~(\Bb) & Paths in  $\Var$-gadgets (vertical) & $\omega^{4}+\Theta_{n,\ell}(\omega^{1})\,>\omega^{4}$  \\ 
		\qquad~(\Bbh)	 & Stairs connecting within $\Var$-gadgets (horizontal) & $\omega^{3}$ \\
		\hline
		Cyan~(\Cc) & Clauses Encoding within $\Var$-gadgets & $\omega^{2}$ \\ 
		\hline
		Green~(\Gg) & (Global) Clause Edges & $\omega^{0}=1$ \\ 
		\hline
	\end{tabular}\smallskip
	\caption{Color-encoding of the weights of the corresponding edges.
		The notation~$\Theta_{n, \ell}(\omega^{1})$ denotes the class of functions~$f$ satisfying~$C_1\cdot \omega^{1} \leq f(\omega^{1}) \leq C_2\cdot \omega^{1}$ for positive constants~$C_1$, $C_2$ that depend on~$n$ and~$\ell$ but are independent of~$\omega$.}
	\label{tab:colors}
\end{table}
Observe that, importantly, all weights used in our construction are bounded by a polynomial in the instance size $|\mathcal{I}|$.

We next introduce auxiliary graphs that serve as building blocks (see~\cref{fig:auxill}) before presenting the full construction of the \textsc{Crossing Number} instance~$(G, k)$.

\begin{figure}[ht]

\subcaptionbox{The variable gadget $\Var^i$, $h=4$\label{fig:var}}
{\begin{tikzpicture}[xscale=0.8,yscale=0.65]\footnotesize
	\tikzonevar{3}{4}{a}
\small
\tikzstyle{every node}=[color=red\blackpart]
\foreach \x in {0,...,4} { \tikzmath{integer \xx; \xx = \x+2;}
	\node[label=left:$r^i_{\xx,L}\,~$] at (l\x a) {};
	\node[label=right:$~\>r^i_{\xx,R}$] at (r\x a) {};
}

\tikzstyle{every node}=[draw, color=blue\blackpart, shape=circle, inner sep=1.1pt, fill=blue\blackpart]

\node[color=blue\blackpart, label=left:$b^i_{1,P}$] at (ll0a) {};
\node[color=blue\blackpart, label=right:$b^i_{1,N}$] at (rr0a) {};
\node[color=blue\blackpart, label=left:$b^i_{6,P}$] at (ll5a) {};
\node[color=blue\blackpart, label=right:$b^i_{6,N}$] at (rr5a) {};

\tikzstyle{every node}=[draw, color=red\blackpart, shape=circle, inner sep=1.1pt, fill=red\blackpart]

\node[label=left:$r^i_{1,L}\!\!$] at (l-1a) {};
\node[label=right:$\!r^i_{1,R}$] at (r-1a) {};
\node[label=left:$r^i_{7,L}\!\!$] at (l5a) {};
\node[label=right:$\!r^i_{7,R}$] at (r5a) {};

\tikzstyle{every node}=[color=blue\blackpart]
\foreach \x in {1,...,4} { \tikzmath{integer \xx; \xx = \x+1;}
	\node[label=left:$b^i_{\xx,P}~~$] at (ll\x a) {};
	\node[label=right:$~~~b^i_{\xx,N}$] at (rr\x a) {};
}

\tikzstyle{every node}=[color=black]
\foreach \x in {1,...,4} { \tikzmath{integer \xx; \xx = \x+1;}
	\node[label=right:$\!\!v^i_{\xx,P}$] at (b\x a) {};
	\node[label=left:$v^i_{\xx,N}\!\!\!$] at (c\x a) {};
}
\node[label=right:$~u^i_0$] at (ada) {};
\node[label=right:$~w^i_0$] at (aua) {};
\node[label=below:$u^i_1$] at (adda) {};
\node[label=above:$w^i_1$] at (auua) {};



\end{tikzpicture}}%
\hfill
\subcaptionbox{The frame with $n$ variable gadgets for $n=3$, $h=4$ \label{fig:frame}}
{\begin{tikzpicture}[xscale=0.5,yscale=0.55]\small
\def\ww{15}\def\hh{8}\def\hho{4}
\tikzstyle{every node}=[draw, varibars, shape=circle, inner sep=1pt, fill]
\tikzstyle{every path}=[draw, framebars, rounded corners=3pt]
\draw (0,0)--(-0.3,-0)--(-0.3,\hh)--(0,\hh) (\ww,0)--(\ww.3,-0)--(\ww.3,\hh)--(\ww,\hh);
\draw (0,-0.03) node{} --(\ww,-0.03) node{} --(\ww,\hh.03) node{} --(0,\hh.03) node{} -- cycle;

\tikzstyle{every node}=[color=red\blackpart, shape=circle, inner sep=1.1pt, fill=red\blackpart]
\tikzstyle{every path}=[draw]

\foreach \x in {0,1,...,\hho} {
	\draw[color=red\blackpart] (0,\x+2) node{} --++ (1,0) (\ww,\x+2) node{} --++ (-1,0);
	\draw[color=red\blackpart] (4,\x+2) --++ (2,0) (9,\x+2) --++ (1,0) (10,\x+2) --++ (1,0);
}

\foreach \v/\a in {2.5/a,7.5/b,12.5/c} {
	\tikzonevar{\v}{\hho}{\a}
}

\tikzstyle{every node}=[color=red\blackpart]
\foreach \x in {0,...,4} { \tikzmath{integer \xx; \xx = \x+2;}
	\node[label=left:$r^0_{\xx,R}$] at (0, \x+2) {};
	\node[label=right:$r^4_{\xx,L}$] at (15, \x+2) {};
}

\tikzstyle{every node}=[color=black, inner sep=1pt]
\node[label=below:{$\!x_0^1=u^1_0\!$}] at (ada) {};
\node[label=above:{$\!x_1^1=w^1_0\!$}] at (aua) {};
\node[label=below:{$\!x_0^2=u^2_0\!$}] at (adb) {};
\node[label=above:{$\!x_1^2=w^2_0\!$}] at (aub) {};
\node[label=below:{$\!x_0^3=u^3_0\!$}] at (adc) {};
\node[label=above:{$\!x_1^3=w^3_0\!$}] at (auc) {};

\node[label=below:{$u^{BL}$}] at (0,-0.03) {};
\node[label=below:{$u^{BR}$}] at (\ww,-0.03) {};
\node[label=above:{$u^{TR}$}] at (\ww,\hh.03) {};
\node[label=above:{$u^{TL}$}] at (0,\hh.03) {};

\end{tikzpicture}\bigskip}
	\caption{Auxiliary graphs. Note that, for each $i\in[n]$, the $8$-tuple $\{u_0^i,u_1^i,r_{1,L}^i,r_{1,R}^i,$ $w_0^i,w_1^i,r_{h+3,L}^i,r_{h+3,R}^i\}$ forms a vertex cut in the frame graph.
	}\label{fig:auxill}
\end{figure}

\subsubsection{Variable gadgets.}\label{par:var}

We begin by defining \emph{$\Var$-gadgets}.
For each~$i\in [n]$, we construct a~$\Var^i$-gadget of height~$h\in\mathbb{Z}_{>0}$ (see an example of~$\Var^i$ for~$h=4$ in~\cref{fig:var}, the precise value of~$h$ will be specified later). 

We first define the vertex set of $\Var^i$ as
$$
V(\Var^i) =\{b_{j, P}^i, b_{j, N}^i, v_{j, P}^i, v_{j, N}^i\}_{j\in [h+2]}\cup \{r_{j, L}^i, r_{j, R}^i\}_{j\in [h+3]}\cup \{w_0^i, u_0^i, w_1^i, u_1^i\}.$$
Vertices ${w_0^i, u_0^i, w_1^i, u_1^i}$ are referred to as the \emph{corner} vertices of $\Var^i$.
We then add six paths:
\begin{itemize}
	\item two \Bb-paths (using \Bb-edges) traverse the vertices $\{b_{j, P}^i\}_{j\in [h+2]}$ and $\{b_{j, N}^i\}_{j\in [h+2]}$, referred to as \texttt{B-pos} and \texttt{B-neg}, respectively;
	\item two \LB-paths traverse $\{v_{j, P}^i\}_{j\in [h+2]}$~(\texttt{LB-pos}) and $\{v_{j, N}^i\}_{j\in [h+2]}$~(\texttt{LB-neg});
	\item two \Rr-paths traverse $\{r_{j, L}^i\}_{j\in [h+3]}$~(\texttt{R-left}) and $\{r_{j, R}^i\}_{j\in [h+3]}$~(\texttt{R-right}).
\end{itemize}

These paths are connected to the corner vertices $w_0^i, u_0^i, w_1^i, u_1^i$ using \HB-edges as follows:
the \texttt{B-pos} and~\texttt{B-neg} paths are connected to $w_0^i$ and $u_0^i$ via their endpoints,
the \texttt{LB-pos} and \texttt{LB-neg} paths are connected analogously to $w_1^i$ and $u_1^i$,
and the \texttt{R-left} and \texttt{R-right} paths connect to all four corner vertices
(see \cref{fig:var}).

Finally, for each $j \in [2, h+1]$, we add horizontal \emph{stairs} between the \texttt{-pos} and \texttt{-neg} paths by connecting \texttt{B-/LB-pos} to \texttt{B-/LB-neg} with \Bbh-edges $b_{j, P}^i v_{j, P}^i$~and~$b_{j, N}^i v_{j, N}^i$, respectively.
The weights of the edges in $\Var^i$ are as specified in \cref{tab:colors};
in particular, for the \Rr-paths, the edges $r_{j,L}^ir_{j+1,L}^i$ and $r_{j,R}^ir_{j+1,R}^i$ have weight exactly $\omega^4 + j(j+1)\omega$,
while for the \Bb-paths, the edges $b_{j,P}^ib_{j+1,P}^i$ and $b_{j,N}^ib_{j+1,N}^i$ have weight exactly $\omega^4 + j(j+2)\omega$.

\subsubsection{The frame.}\label{sec:frame}

We construct the \emph{frame} for $n$ $\Var$-gadgets of height $h$, where $n, h\in\mathbb{Z}_{>0}$.

The construction begins with a \HB-cycle on four vertices, $u^{BL}$ (bottom-left), $u^{TL}$ (top-left), $u^{TR}$ (top-right), and $u^{BR}$ (bottom-right), connected in the specified order using \HB-edges. The edge between $u^{BL}$ and $u^{BR}$ is then subdivided~$n$ times by adding vertices~$\{x_0^i\}_{i\in[n]}$, and analogously, the edge between~$u^{TL}$ and~$u^{TR}$ is subdivided by adding vertices $\{x_1^i\}_{i\in[n]}$.

Next, we add another \HB-edge between~$u^{BL}$ and~$u^{TL}$ and subdivide it $(h+1)$ times by adding vertices~$\{r_{j, R}^0\}_{j\in[2, h+2]}$, and similarly, we add an \HB-edge between~$u^{TR}$ and $u^{BR}$ and subdivide it $(h+1)$ times by adding vertices~$\{r_{j, L}^{n+1}\}_{j\in[2, h+2]}$. The resulting graph of this construction (see \cref{fig:frame}) is called the \emph{frame}~$F$.

\subsubsection{The frame with variable gadgets.}\label{sec:framev}
Finally, we attach~$n$~$\Var$-gadgets to the frame~$F$. 
For each $i\in [n]$, we introduce a~$\Var^i$-gadget as described in~\cref{par:var}, and we identify the vertices~$u_0^i$ and~$w_0^i$ of~$\Var^i$ with the frame vertices~$x_0^i$ and~$x_1^i$, respectively.
We then add horizontal \emph{stairs} connecting the \Rr-paths of neighboring~$\Var$-gadgets and the frame: for each~$i\in [n+1]$ and~$j\in [2, h+2]$, we add an \Rrh-edge~$r_{j, R}^{i-1}r_{j, L}^{i}$.
More concretely, for each~$i\in[n+1]$, we connect the \texttt{\Rr-right} path of~$\Var^{i-1}$ (or, if~$i=1$, a subdivision of the frame side~$u^{TL}u^{BL}$) to the \texttt{\Rr-left} path of~$\Var^{i}$ (or, if~$i=n+1$, a subdivision of the frame side~$u^{TR}u^{BR}$).
All new edges receive weights as specified in~\cref{tab:colors}.

This completes the construction of the {\em frame with variable gadgets}, that is, the graph~$G'$ (see \cref{fig:frame} for an illustration with~$h=4$ and~$n=3$).
Note that $G'$ does not yet include the clause edges (see~\cref{fig:example-noflip}) or the interpretation of variable occurrences within clauses (the cells in~\cref{fig:cells}).

For simplicity, we refer to \cref{fig:frame} to illustrate the graph $G'$ constructed so far.
Note the natural interpretation of the \texttt{R-left} and \texttt{R-right} paths within each gadget $\Var^i$: to facilitate connections with~$\Var^{i-1}$ and~$\Var^{i+1}$, the \texttt{R-left} path is naturally drawn to the left of \texttt{R-right}.
In contrast, the \texttt{B-/LB-pos} and~\texttt{B-/LB-neg} paths of $\Var^i$ are symmetric and not adjacent outside of the gadget $\Var^i$.
Consequently, these paths can be drawn flexibly, with \texttt{B-/LB-pos} placed either to the left or to the right of \texttt{B-/LB-neg};
this flexibility will later encode the truth value of the variable represented by $\Var^i$.

\subsection{The Full Reduction}\label{sub:fullred}

Consider an instance~$(\mathcal{C}, \mathcal{V})$ of \textsc{Satisfiability}, where~$|\mathcal{C}|=\ell$ and~$|\mathcal{V}|=n$.
We construct a corresponding instance~$(G, k)$ of \textsc{Crossing Number} as described below. 
A schematic representation of the construction is shown in~\cref{fig:example-noflip}.

\begin{figure}[ht]
	\centerline{\scalebox{1}{\def\slightupendx{--++(1.25,0) --++(1.5,.5) -- ++(1.5,0)}

\begin{tikzpicture}[xscale=0.5,yscale=0.475]\small
\def\ww{26}\def\hh{19}\def\hho{15}
\tikzstyle{every node}=[draw, varibars, shape=circle, inner sep=1.1pt, fill]
\tikzstyle{every path}=[draw, framebars, rounded corners=3pt]
\draw (0,0)--(-0.3,-0)--(-0.3,\hh)--(0,\hh) (\ww,0)--(\ww.3,-0)--(\ww.3,\hh)--(\ww,\hh);
\draw (0,-0.03) node{} --(\ww,-0.03) node{} --(\ww,\hh.03) node{} --(0,\hh.03) node{} -- cycle;

\foreach \v/\a in {3/a,8/b,13/c,18/d,23/f} {
	\tikzonevar{\v}{\hho}{\a}
}
\tikzstyle{every node}=[color=red\blackpart, shape=circle, inner sep=1.1pt, fill=red\blackpart]
\tikzstyle{every path}=[draw]
\foreach \x in {0,1,...,\hho} {
	\draw[color=red\blackpart] (0,\x+2) node{} --(l\x a) (\ww,\x+2) node{} --(r\x f);
	\draw[color=red\blackpart] (r\x a)--(l\x b) (r\x b)--(l\x c) (r\x c)--(l\x d) (r\x d)--++(2,0) (l\x f)--++(-1,0);
}
\tikzstyle{every node}=[draw, color=black, shape=circle, inner sep=0.8pt, fill=black]
\tikzstyle{every path}=[draw, color=cyan\blackpart]
\tikzstyle{blbars}=[black,semithick]

\draw[blbars] (cda)--(b1a)--(c1a) (b3a)--(c3a);
\draw (b1a)--(c2a) --(b2a) --(c3a);
\draw[blbars] (c3a)--(b4a)--(c4a) (c4a)--(b5a)--(c5a) (b7a)--(c7a);
\draw (c5a)--(b6a)--(c6a)--(b7a); 
\draw[blbars] (c7a)--(b8a)--(c8a) (c8a)--(b9a)--(c9a) (b10a)--(c10a) (b11a)--(c11a);
\draw (b9a)--(c10a)--(b11a); 
\draw[blbars] (bua)--(c15a)--(b15a)--(c14a)--(b14a)--(c13a)--(b13a)--(c12a)--(b12a)--(c11a);

\draw[blbars] (cdb)--(b1b)--(c1b)--(b2b)--(c2b) (b4b)--(c4b);
\draw (c2b)--(b3b)--(c3b)--(b4b);
\draw[blbars] (c4b)--(b5b)--(c5b) (c5b)--(b6b)--(c6b) (b7b)--(c7b) (b8b)--(c8b);
\draw (b6b)--(c7b)--(b8b); 
\draw[blbars] (c8b)--(b9b)--(c9b) (c9b)--(b10b)--(c10b) (b12b)--(c12b);
\draw (b10b)--(c11b)--(b11b)--(c12b);
\draw[blbars] (bub)--(c15b)--(b15b)--(c14b)--(b14b)--(c13b)--(b13b)--(c12b);

\draw[blbars] (cdc)--(b1c)--(c1c)--(b2c)--(c2c)--(b3c)--(c3c) (b4c)--(c4c) (b5c)--(c5c);
\draw (b3c)--(c4c)--(b5c);
\draw[blbars] (c5c)--(b6c)--(c6c) (c6c)--(b7c)--(c7c) (c9c)--(b9c);
\draw (c7c)--(b8c)--(c8c)--(b9c);
\draw[blbars] (c9c)--(b10c)--(c10c) (c10c)--(b11c)--(c11c) (b13c)--(c13c);
\draw (b11c)--(c12c)--(b12c)--(c13c); 
\draw[blbars] (buc)--(c15c)--(b15c)--(c14c)--(b14c)--(c13c);

\draw[blbars] (cdd)--(b1d)--(c1d)--(b2d)--(c2d)--(b3d)--(c3d)--(b4d)--(c4d) (b6d)--(c6d);
\draw (b4d)--(c5d)--(b5d)--(c6d);
\draw[blbars] (c6d)--(b7d)--(c7d) (c7d)--(b8d)--(c8d) (b9d)--(c9d) (b10d)--(c10d);
\draw (b8d)--(c9d)--(b10d);
\draw[blbars] (c10d)--(b11d)--(c11d) (c11d)--(b12d)--(c12d) (b14d)--(c14d);
\draw (c12d)--(b13d)--(c13d)--(b14d); 
\draw[blbars] (bud)--(c15d)--(b15d)--(c14d);

\draw[blbars] (cdf)--(b1f)--(c1f)--(b2f)--(c2f)--(b3f)--(c3f)--(b4f)--(c4f)--(b5f)--(c5f) (b7f)--(c7f);
\draw (c5f)--(b6f)--(c6f)--(b7f);
\draw[blbars] (c7f)--(b8f)--(c8f) (c8f)--(b9f)--(c9f) (b11f)--(c11f);
\draw (b9f)--(c10f)--(b10f)--(c11f);
\draw[blbars] (c11f)--(b12f)--(c12f) (c12f)--(b13f)--(c13f) (b14f)--(c14f) (b15f)--(c15f);
\draw (b15f)--(c14f)--(b13f);
\draw[blbars] (buf) -- ++(2,-.5);

\tikzstyle{every path}=[draw, color=green\gblackpart, rounded corners=4pt]
\tikzstyle{every node}=[color=green\gblackpart, shape=circle, inner sep=1.1pt, fill=green\gblackpart]
\def\allclauses{%
\draw (0,2.75) node[label=left:$c_{1,L}~$] {} --++(26,5.75) node[label=right:$~c_{1,R}$] {};
\draw (0,6.75) node[label=left:$c_{2,L}~$] {} --++(26,5.75)  node[label=right:$~c_{2,R}$] {};
\draw (0,10.75) node[label=left:$c_{3,L}~$] {} --++(26,5.75)  node[label=right:$~c_{3,R}$] {};
}\allclauses  

\begin{scope}[on background layer]
\tikzstyle{every path}=[draw=none, fill=gray!20!white]
\foreach \i / \x in {1/a,2/b,3/c,4/d,5/f} { \foreach \zz in {0,4,...,8} {
	\tikzmath {integer \z; \z = \zz+\i;}
 	\path (ll\z\x) rectangle ++(4,2);
 	\path (ll\z\x)++(0,-0.5) rectangle ++(-1,2);
	\ifnum \i >2
 	\path (ll\z\x)++(4,0.5) rectangle ++(1,2);
	\fi
}}
\tikzstyle{every path}=[draw, color=white, line width=2.5pt, rounded corners=4pt]
\tikzstyle{every node}=[draw=none]
\allclauses
\end{scope}

\tikzstyle{every node}=[color=black, inner sep=1pt]
\node[label=above:{$\!x_1\!$}] at (aua) {};
\node[label=above:{$\!x_2\!$}] at (aub) {};
\node[label=above:{$\!x_3\!$}] at (auc) {};
\node[label=above:{$\!x_4\!$}] at (aud) {};
\node[label=above:{$\!x_5\!$}] at (auf) {};

\end{tikzpicture}}}
	\caption{For an example, consider an instance of \textsc{Satisfiability} with variables $\mathcal{V}=\{x_1,x_2,x_3,x_4,x_5\}$ and clauses $\mathcal{C}=\{(x_1 \vee \overline{x_2} \vee x_4 \vee \overline{x_5}),(\overline{x_1} \vee \overline{x_3} \vee x_5),(x_2 \vee x_3 \vee \overline{x_4})\}.$
		The corresponding graph~$G$ is constructed as the instance of \textsc{Crossing Number} produced by the reduction.
		In particular, note the addition of the clause edges, drawn in green from left to right across the frame, and the shaded regions indicating the areas through which the clause edges are intended to be routed.}
	\label{fig:example-noflip}
\end{figure}

We start with the graph $G'$, consisting of the frame with $n$ variable $\Var$-gadgets of height~$h=4\ell+n-2$.
Then, for each $i \in [n]$, we encode the occurrences of the variable~$x_i$ in the clauses $C_j$ of $\mathcal{C}$.
Specifically, for each~$j\in[\ell]$, we insert a \emph{cell of~$C_j$} between the \texttt{LB-pos} and \texttt{LB-neg} paths of the gadget~$\Var^i$.
Each of the~$\ell$ cells is formed by two horizontal \LB-edges and three internal edges, whose configuration depends on the type of the cell: $C_{pos}$ if $x_i \in C_j$, $C_{neg}$ if $\overline{x_i} \in C_j$, and $C_{neut}$ if neither $x_i$ nor $\overline{x_i}$ occurs in $C_j$ (see \cref{fig:cells} for an illustration of the three types).
Cells within the same $\Var$-gadget are separated by additional \LB-edges, as depicted in \cref{fig:example-noflip}. 	 

\begin{figure}[ht]
	\centerline{\scalebox{1}{\def\blackpart{!75!black}
\tikzstyle{every node}=[draw, color=black, shape=circle, inner sep=0.8pt, fill=black]
\tikzstyle{every path}=[draw, color=cyan\blackpart]
\tikzstyle{blbars}=[black,semithick]
\begin{subfigure}[b]{0.33\textwidth}
	\centering
\begin{tikzpicture}[scale=0.65]\footnotesize
	\draw (3,0) --++(2,1) node{} --++(-2,0) node{} node{} --++(2,1) node{};
	\draw[blbars] (3,0) node{} --++(2,0) node{} --++(0,2) node{}  --++(-2,0) node{}  -- cycle
	(3,-0.3) -- (3,2.3) (5,-0.3) -- (5,2.3);
	\node[draw=none,fill=none] at (1.7,1) {\texttt{LB-pos}};
	\node[draw=none,fill=none] at (6.3,1) {\texttt{LB-neg}};
	\end{tikzpicture}
	\caption{$C_{\text{pos}}$: $x\in C$} \label{fig:var_pos}
\end{subfigure}
\begin{subfigure}[b]{0.33\textwidth}
	\centering
	\begin{tikzpicture}[scale=0.65]\small
	\draw (8,0) --++(-2,1) node{} --++(2,0) node{} node{} --++(-2,1) node{};
	\draw[blbars] (6,0) node{} --++(2,0) node{} --++(0,2) node{}  --++(-2,0) node{}  -- cycle	
	(6,-0.3) -- (6,2.3) (8,-0.3) -- (8,2.3);
	\node[draw=none,fill=none] at (4.7,1) {\texttt{LB-pos}};
	\node[draw=none,fill=none] at (9.3,1) {\texttt{LB-neg}};
	\end{tikzpicture}
	\caption{$C_{\text{neg}}$: $\overline{x}\in C$} \label{fig:var_neg} 
\end{subfigure}
\begin{subfigure}[b]{0.33\textwidth}
	\centering
	\begin{tikzpicture}[scale=0.65]\small
	\draw (0,0)-- (2, 1) (2,1) -- (0, 2);
	\draw[blbars] (0,0) node{} --++(2,0) node{} --++(0,2) node{}  --++(-2,0) node{}  -- cycle
	(0,1) node{} -- (2,1) node{}
	(0,-0.3) -- (0,2.3) (2,-0.3) -- (2,2.3);
	\node[draw=none,fill=none] at (-1.3,1) {\texttt{LB-pos}};
	\node[draw=none,fill=none] at (3.3,1) {\texttt{LB-neg}};
	\end{tikzpicture}
	\caption{$C_{\text{neut}}$: neither $x$ nor $\overline{x}$ is in $C$ }\label{fig:var_neut}
\end{subfigure}}}
	\caption{Cell types; for cases of variable $x$ occurrence in clause $C$.}
	\label{fig:cells}
\end{figure}

Formally, for each $i \in [n]$, we turn the gadget $\Var^i$ of $G'$ into the \emph{loaded variable gadget}~$\Var_+^i$ of~$G$ as follows.
We begin by adding \LB-edges inside each $\Var$-gadget to separate the cells.
For each $i \in [n]$, we add an \LB-path from $v_{1,N}^i$ to $v_{1+i,P}^i$ running below all cells of the $\Var^i$-gadget along the vertices $v_{1,N}^i$, $v_{2,P}^i$, $v_{2,N}^i$, \ldots, $v_{i,N}^i$, $v_{1+i,P}^i$ in order.
Another \LB-path connects $v_{4\ell+i-1,N}^i$ to $v_{h+2,P}^i$ above all cells, traversing vertices $v_{4\ell+i-1,N}^i$, $v_{4\ell+i,P}^i$, $v_{4\ell+i,N}^i$, \dots, $v_{h+1,N}^i$, $v_{h+2,P}^i$.
Additionally, for each $j\in [\ell-1]$, we add an \LB-path connecting $v_{4j+i-1,N}^i$, $v_{4j+i,P}^i$, $v_{4j+i,N}^i$, $v_{4j+i+1,P}^i$ between consecutive cells.

After that, we insert the cells themselves in bottom-up order.
For each $j\in[\ell]$, we first add two \LB-edges $v_{4j+i-3,P}^i v_{4j+i-3,N}^i$ and $v_{4j+i-1,P}^i v_{4j+i-1,N}^i$.
Then, we encode the clause~$C_j$ in its cell in $\Var_+^i$ as follows: 
\begin{itemize}
	\item[-] if $x_i\in C_j$, we insert a $C_{pos}$ cell (\cref{fig:var_pos}) by adding three \Cc-edges along the path $v_{4j+i-3,P}^i$, $v_{4j+i-2,N}^i$, $v_{4j+i-2,P}^i$, $v_{4j+i-1,N}^i$;
	\item[-] if $\overline{x_i}\in C_j$, we insert a $C_{neg}$ cell (\cref{fig:var_neg}) along $v_{4j+i-3,N}^i$, $v_{4j+i-2,P}^i$, $v_{4j+i-2,N}^i$, $v_{4j+i-1,P}^i$;
	\item[-] and if neither $x_i$ nor $\overline{x_i}$ occurs in $C_j$, we insert a $C_{neut}$ cell (\cref{fig:var_neut}), consisting of one \LB-edge $v_{4j+i-2,P}^i v_{4j+i-2,N}^i$ and two \Cc-edges $v_{4j+i-3,P}^i v_{4j+i-2,N}^i$ and $v_{4j+i-2,N}^i v_{4j+i-1,P}^i$.
\end{itemize}

Finally, for each clause $C_j$, we introduce in~$G$ a \Gg-edge representing the clause itself.
We subdivide two vertical \HB-edges of the frame and connect the new vertices to form the clause edge: 
specifically, we subdivide the edge between $r^0_{4j-2, R}$ and $r^0_{4j-1, R}$ on the left vertical side $u^{BL} u^{TL}$ with a new vertex~$c_{j,L}$,
and the edge between $r^{n+1}_{4j+n-1, L}$, $r^{n+1}_{4j+n, L}$ on the right vertical side $u^{BR} u^{TR}$ with a new vertex~$c_{j,R}$,
and add the~\Gg-edge~$c_{j,L}c_{j,R}$. See \cref{fig:example-noflip} for an illustration.
Note that the indices of the subdivided edges are shifted by~$n+1$ from left to right to maintain alignment with the variable gadgets.

This concludes the construction of the graph $G$.
The reduction outputs the pair $(G, k)$ as an instance of \textsc{Crossing Number}, where, for $h=4\ell+n-2$, the parameter $k$ is given by
\begin{equation}\label{eq:kvalue}
k=2n(2h+1)\omega^7+2n\ell\omega^6+4n\ell\omega^4+2n\sum\limits_{j=2}^{h+1}j(j+1)\omega^4+
2n\sum\limits_{j=1}^{h+1}j(j+2)\omega^4+ n\ell\omega^2+ (\omega^2-1).
\end{equation}

\begin{figure}[ht]
	\centerline{\scalebox{1}{\def\slightupendx{--++(1.25,0) --++(1.5,.5) -- ++(1.5,0)}

\begin{tikzpicture}[xscale=0.5,yscale=0.55]\small
\def\ww{26}\def\hh{19}\def\hho{15}
\tikzstyle{every node}=[draw, varibars, shape=circle, inner sep=1.1pt, fill]
\tikzstyle{every path}=[draw, framebars, rounded corners=3pt]
\draw (0,0)--(-0.3,-0)--(-0.3,\hh)--(0,\hh) (\ww,0)--(\ww.3,-0)--(\ww.3,\hh)--(\ww,\hh);
\draw (0,-0.03) node{} --(\ww,-0.03) node{} --(\ww,\hh.03) node{} --(0,\hh.03) node{} -- cycle;

\foreach \v/\a in {3/a,8/b,13/c,18/d,23/f} {
	\tikzonevar{\v}{\hho}{\a}
}
\tikzstyle{every node}=[color=red\blackpart, shape=circle, inner sep=1.1pt, fill=red\blackpart]
\tikzstyle{every path}=[draw]
\foreach \x in {0,1,...,\hho} {
	\draw[color=red\blackpart] (0,\x+2) node{} --(l\x a) (\ww,\x+2) node{} --(r\x f);
	\draw[color=red\blackpart] (r\x a)--(l\x b) (r\x b)--(l\x c) (r\x c)--(l\x d) (r\x d)--++(2,0) (l\x f)--++(-1,0);
}
\tikzstyle{every node}=[draw, color=black, shape=circle, inner sep=0.8pt, fill=black]
\tikzstyle{every path}=[draw, color=cyan\blackpart]
\tikzstyle{blbars}=[black,semithick]

\draw[blbars] (cda)--(b1a)--(c1a) (b3a)--(c3a);
\draw (b1a)--(c2a) --(b2a) --(c3a);
\draw[blbars] (c3a)--(b4a)--(c4a) (c4a)--(b5a)--(c5a) (b7a)--(c7a);
\draw (c5a)--(b6a)--(c6a)--(b7a); 
\draw[blbars] (c7a)--(b8a)--(c8a) (c8a)--(b9a)--(c9a) (b10a)--(c10a) (b11a)--(c11a);
\draw (b9a)--(c10a)--(b11a); 
\draw[blbars] (bua)--(c15a)--(b15a)--(c14a)--(b14a)--(c13a)--(b13a)--(c12a)--(b12a)--(c11a);

\draw[blbars] (cdb)--(b1b)--(c1b)--(b2b)--(c2b) (b4b)--(c4b);
\draw (c2b)--(b3b)--(c3b)--(b4b);
\draw[blbars] (c4b)--(b5b)--(c5b) (c5b)--(b6b)--(c6b) (b7b)--(c7b) (b8b)--(c8b);
\draw (b6b)--(c7b)--(b8b); 
\draw[blbars] (c8b)--(b9b)--(c9b) (c9b)--(b10b)--(c10b) (b12b)--(c12b);
\draw (b10b)--(c11b)--(b11b)--(c12b);
\draw[blbars] (bub)--(c15b)--(b15b)--(c14b)--(b14b)--(c13b)--(b13b)--(c12b);

\draw[blbars] (bdc)--(c1c)--(b1c)--(c2c)--(b2c)--(c3c)--(b3c) (b4c)--(c4c) (b5c)--(c5c);
\draw (c3c)--(b4c)--(c5c);
\draw[blbars] (b5c)--(c6c)--(b6c) (b6c)--(c7c)--(b7c) (c9c)--(b9c);
\draw (b7c)--(c8c)--(b8c)--(c9c);
\draw[blbars] (b9c)--(c10c)--(b10c) (b10c)--(c11c)--(b11c) (b13c)--(c13c);
\draw (c11c)--(b12c)--(c12c)--(b13c); 
\draw[blbars] (cuc)--(b15c)--(c15c)--(b14c)--(c14c)--(b13c);

\draw[blbars] (bdd)--(c1d)--(b1d)--(c2d)--(b2d)--(c3d)--(b3d)--(c4d)--(b4d) (b6d)--(c6d);
\draw (c4d)--(b5d)--(c5d)--(b6d);
\draw[blbars] (b6d)--(c7d)--(b7d) (b7d)--(c8d)--(b8d) (b9d)--(c9d) (b10d)--(c10d);
\draw (c8d)--(b9d)--(c10d);
\draw[blbars] (b10d)--(c11d)--(b11d) (b11d)--(c12d)--(b12d) (b14d)--(c14d);
\draw (b12d)--(c13d)--(b13d)--(c14d); 
\draw[blbars] (cud)--(b15d)--(c15d)--(b14d);

\draw[blbars] (bdf)--(c1f)--(b1f)--(c2f)--(b2f)--(c3f)--(b3f)--(c4f)--(b4f)--(c5f)--(b5f) (b7f)--(c7f);
\draw (b5f)--(c6f)--(b6f)--(c7f);
\draw[blbars] (b7f)--(c8f)--(b8f) (b8f)--(c9f)--(b9f) (b11f)--(c11f);
\draw (c9f)--(b10f)--(c10f)--(b11f);
\draw[blbars] (b11f)--(c12f)--(b12f) (b12f)--(c13f)--(b13f) (b14f)--(c14f) (b15f)--(c15f);
\draw (c15f)--(b14f)--(c13f);
\draw[blbars] (cuf) -- ++(-2,-.5);

\tikzstyle{every path}=[draw, color=green\gblackpart, rounded corners=4pt]
\tikzstyle{every node}=[color=green\gblackpart, shape=circle, inner sep=1.1pt, fill=green\gblackpart]
\def\allclauses{%
\draw (0,2.75) node[label=left:$C_1~$] {} --++(1.75,0) \slightup \slightup \slightup \slightup \jumponeupend node{};
\draw (0,6.75) node[label=left:$C_2~$] {} --++(1.75,0) \slightup \slightup \jumponeup \slightup \slightupendx node{};
\draw (0,10.75) node[label=left:$C_3~$] {} --++(1.75,0) \slightup \jumponeup \slightup \slightup \slightupendx node{};
}\allclauses  

\begin{scope}[on background layer]
\tikzstyle{every path}=[draw=none, fill=gray!20!white]
\foreach \i / \x in {1/a,2/b,3/c,4/d,5/f} { \foreach \zz in {0,4,...,8} {
	\tikzmath {integer \z; \z = \zz+\i;}
 	\path (ll\z\x) rectangle ++(4,2);
 	\path (ll\z\x)++(0,-0.5) rectangle ++(-1,2);
	\ifnum \i >2
 	\path (ll\z\x)++(4,0.5) rectangle ++(1,2);
	\fi
}}
\tikzstyle{every path}=[draw, color=white, line width=2.5pt, rounded corners=4pt]
\tikzstyle{every node}=[draw=none]
\allclauses
\end{scope}

\tikzstyle{every node}=[color=black, inner sep=1pt]
\node[label=above:{$\!x_1=\true\!$}] at (aua) {};
\node[label=above:{$\!x_2=\true\!$}] at (aub) {};
\node[label=above:{$\!x_3=\false\!$}] at (auc) {};
\node[label=above:{$\!x_4=\false\!$}] at (aud) {};
\node[label=above:{$\!x_5=\false\!$}] at (auf) {};

\end{tikzpicture}}}
	\caption{A drawing of the graph $G$ from \cref{fig:example-noflip}, constructed from the \textsc{Satisfiability} instance given by $\mathcal{V}=\{x_1,x_2,x_3,x_4,x_5\}$ and $\mathcal{C}=\{(x_1 \vee \overline{x_2} \vee x_4 \vee \overline{x_5}),(\overline{x_1} \vee \overline{x_3} \vee x_5),(x_2 \vee x_3 \vee \overline{x_4})\}$.
		\newline
		The drawing $\mathcal{G}$ of $G$ depicted in the figure corresponds to the satisfying assignment $x_1=x_2=\true$, $x_3=x_4=x_5=\false$.
		Clause $C_1=(x_1 \vee \overline{x_2} \vee x_4 \vee \overline{x_5})$ is satisfied by the variable $x_5$; note that the \Gg-edge of $C_1$ makes an extra jump-up in the drawing area of $\Var^5$, yet crossing only one \Cc-edge there, as in other gadgets. Clause
		$C_2=(\overline{x_1} \vee \overline{x_3} \vee x_5)$ is satisfied by~$x_3$ and $C_3=(x_2 \vee x_3 \vee \overline{x_4})$ is satisfied by $x_2$.
		For comparison, \cref{fig:example-assignments-b} illustrates a drawing corresponding to an unsatisfying assignment.
	}\label{fig:example-assignments}\label{fig:example-assignments-a}
\end{figure}
\begin{figure}[ht]
	\centerline{\scalebox{1}{\def\slightupendx{--++(1.25,0) --++(1.5,.5) -- ++(1.5,0)}

\begin{tikzpicture}[xscale=0.5,yscale=0.55]\small
\def\ww{26}\def\hh{19}\def\hho{15}
\tikzstyle{every node}=[draw, varibars, shape=circle, inner sep=1.1pt, fill]
\tikzstyle{every path}=[draw, framebars, rounded corners=3pt]
\draw (0,0)--(-0.3,-0)--(-0.3,\hh)--(0,\hh) (\ww,0)--(\ww.3,-0)--(\ww.3,\hh)--(\ww,\hh);
\draw (0,-0.03) node{} --(\ww,-0.03) node{} --(\ww,\hh.03) node{} --(0,\hh.03) node{} -- cycle;

\foreach \v/\a in {3/a,8/b,13/c,18/d,23/f} {
	\tikzonevar{\v}{\hho}{\a}
}
\tikzstyle{every node}=[color=red\blackpart, shape=circle, inner sep=1.1pt, fill=red\blackpart]
\tikzstyle{every path}=[draw]
\foreach \x in {0,1,...,\hho} {
	\draw[color=red\blackpart] (0,\x+2) node{} --(l\x a) (\ww,\x+2) node{} --(r\x f);
	\draw[color=red\blackpart] (r\x a)--(l\x b) (r\x b)--(l\x c) (r\x c)--(l\x d) (r\x d)--++(2,0) (l\x f)--++(-1,0);
}
\tikzstyle{every node}=[draw, color=black, shape=circle, inner sep=0.8pt, fill=black]
\tikzstyle{every path}=[draw, color=cyan\blackpart]
\tikzstyle{blbars}=[black,semithick]

\draw[blbars] (cda)--(b1a)--(c1a) (b3a)--(c3a);
\draw (b1a)--(c2a) --(b2a) --(c3a);
\draw[blbars] (c3a)--(b4a)--(c4a) (c4a)--(b5a)--(c5a) (b7a)--(c7a);
\draw (c5a)--(b6a)--(c6a)--(b7a); 
\draw[blbars] (c7a)--(b8a)--(c8a) (c8a)--(b9a)--(c9a) (b10a)--(c10a) (b11a)--(c11a);
\draw (b9a)--(c10a)--(b11a); 
\draw[blbars] (bua)--(c15a)--(b15a)--(c14a)--(b14a)--(c13a)--(b13a)--(c12a)--(b12a)--(c11a);

\draw[blbars] (cdb)--(b1b)--(c1b)--(b2b)--(c2b) (b4b)--(c4b);
\draw (c2b)--(b3b)--(c3b)--(b4b);
\draw[blbars] (c4b)--(b5b)--(c5b) (c5b)--(b6b)--(c6b) (b7b)--(c7b) (b8b)--(c8b);
\draw (b6b)--(c7b)--(b8b); 
\draw[blbars] (c8b)--(b9b)--(c9b) (c9b)--(b10b)--(c10b) (b12b)--(c12b);
\draw (b10b)--(c11b)--(b11b)--(c12b);
\draw[blbars] (bub)--(c15b)--(b15b)--(c14b)--(b14b)--(c13b)--(b13b)--(c12b);

\draw[blbars] (cdc)--(b1c)--(c1c)--(b2c)--(c2c)--(b3c)--(c3c) (c4c)--(b4c) (c5c)--(b5c);
\draw (b3c)--(c4c)--(b5c);
\draw[blbars] (c5c)--(b6c)--(c6c) (c6c)--(b7c)--(c7c) (c9c)--(b9c);
\draw (c7c)--(b8c)--(c8c)--(b9c);
\draw[blbars] (c9c)--(b10c)--(c10c) (c10c)--(b11c)--(c11c) (c13c)--(b13c);
\draw (b11c)--(c12c)--(b12c)--(c13c); 
\draw[blbars] (buc)--(c15c)--(b15c)--(c14c)--(b14c)--(c13c);

\draw[blbars] (bdd)--(c1d)--(b1d)--(c2d)--(b2d)--(c3d)--(b3d)--(c4d)--(b4d) (b6d)--(c6d);
\draw (c4d)--(b5d)--(c5d)--(b6d);
\draw[blbars] (b6d)--(c7d)--(b7d) (b7d)--(c8d)--(b8d) (b9d)--(c9d) (b10d)--(c10d);
\draw (c8d)--(b9d)--(c10d);
\draw[blbars] (b10d)--(c11d)--(b11d) (b11d)--(c12d)--(b12d) (b14d)--(c14d);
\draw (b12d)--(c13d)--(b13d)--(c14d); 
\draw[blbars] (cud)--(b15d)--(c15d)--(b14d);

\draw[blbars] (bdf)--(c1f)--(b1f)--(c2f)--(b2f)--(c3f)--(b3f)--(c4f)--(b4f)--(c5f)--(b5f) (b7f)--(c7f);
\draw (b5f)--(c6f)--(b6f)--(c7f);
\draw[blbars] (b7f)--(c8f)--(b8f) (b8f)--(c9f)--(b9f) (b11f)--(c11f);
\draw (c9f)--(b10f)--(c10f)--(b11f);
\draw[blbars] (b11f)--(c12f)--(b12f) (b12f)--(c13f)--(b13f) (b14f)--(c14f) (b15f)--(c15f);
\draw (c15f)--(b14f)--(c13f);
\draw[blbars] (cuf) -- ++(-2,-.5);

\tikzstyle{every path}=[draw, color=green\gblackpart, rounded corners=4pt]
\tikzstyle{every node}=[color=green\gblackpart, shape=circle, inner sep=1.1pt, fill=green\gblackpart]
\def\allclauses{%
\draw (0,2.75) node[label=left:$C_1~$] {} --++(1.75,0) \slightup \slightup \slightup \slightup \jumponeupend node{};
\draw (0,6.75) node[label=left:$C_2~$] {} --++(1.75,0) \slightup \slightup \jumponeup \slightup \slightupendx node{};
\draw (0,10.75) node[label=left:$C_3~$] {} --++(1.75,0) \slightup \jumponeup \slightup \slightup \slightupendx node{};
}\allclauses  

\begin{scope}[on background layer]
\tikzstyle{every path}=[draw=none, fill=gray!20!white]
\foreach \i / \x in {1/a,2/b,3/c,4/d,5/f} { \foreach \zz in {0,4,...,8} {
	\tikzmath {integer \z; \z = \zz+\i;}
 	\path (ll\z\x) rectangle ++(4,2);
 	\path (ll\z\x)++(0,-0.5) rectangle ++(-1,2);
	\ifnum \i >2
 	\path (ll\z\x)++(4,0.5) rectangle ++(1,2);
	\fi
}}
\tikzstyle{every path}=[draw, color=white, line width=2.5pt, rounded corners=4pt]
\tikzstyle{every node}=[draw=none]
\allclauses
\end{scope}

\tikzstyle{every node}=[color=black, inner sep=1pt]
\node[label=above:{$\!x_1=\true\!$}] at (aua) {};
\node[label=above:{$\!x_2=\true\!$}] at (aub) {};
\node[label=above:{$\!x_3=\true\!$}] at (auc) {};
\node[label=above:{$\!x_4=\false\!$}] at (aud) {};
\node[label=above:{$\!x_5=\false\!$}] at (auf) {};

\end{tikzpicture}}}
	\caption{Another drawing of the graph $G$ from \cref{fig:example-noflip}, constructed from the \textsc{Satisfiability} instance given by $\mathcal{V}=\{x_1,x_2,x_3,x_4,x_5\}$ and $\mathcal{C}=\{(x_1 \vee \overline{x_2} \vee x_4 \vee \overline{x_5}),(\overline{x_1} \vee \overline{x_3} \vee x_5),(x_2 \vee x_3 \vee \overline{x_4})\}$,
		shown for comparison with the drawing in \cref{fig:example-assignments-a}.
		\newline
		The depicted drawing $\mathcal{G}'$ of $G$ corresponds to the unsatisfying assignment $x_1=x_2=x_3=\true$, $x_4=x_5=\false$.
		Clause $C_1=(x_1 \vee \overline{x_2} \vee x_4 \vee \overline{x_5})$ is satisfied by the variable $x_5$, and $C_3=(x_2 \vee x_3 \vee \overline{x_4})$ is satisfied by~$x_2$, as in~\cref{fig:example-assignments-a}.
		The unsatisfied clause $C_2=(\overline{x_1} \vee \overline{x_3} \vee x_5)$ differs in that the \Gg-edge of $C_2$ cannot make the required extra jump-up without crossing more than one \Cc-edge inside one of $\Var$-gadgets (or other heavier edges).
		As a result, the \Gg-edge of $C_2$ crosses two \Cc-edges within the drawing area of $\Var^3\!$, which unavoidably increases the number of crossings, yielding $\crn(\mathcal{G}')>k$.
	}\label{fig:example-assignments-b}
	
\end{figure}

\subsection{Drawings Claims}
\label{subsec:claims}
Up to this point, all figures have been presented purely for illustrative purposes, without a formal justification of why particular drawings of the constructed graphs are optimal or even admissible.
This subsection is devoted to clarifying the structural constraints that any drawing $\mathcal{G}$ of the constructed instance~$(G,k)$ of \textsc{Crossing Number} must satisfy in order to be a valid solution, that is, in order to have at most~$k$ crossings.

We therefore consider the instance $(G,k)$ of \textsc{Crossing Number} produced by the reduction described in \cref{sub:fullred}.
Following the construction of the graph $G$, we derive a sequence of observations and claims that must hold for every drawing of $G$ whose crossing number does not exceed $k$.
We begin with constraints imposed by the edges of largest weight, focusing first on the frame and the $\Var$-gadgets.
Subsequently, we analyze the implications of these constraints for the gadgets encoding the clauses.

\begin{obs}\label{obs:HB}
	If a drawing $\calG$ of $G$ is a solution of the instance $(G, k)$, then no \HB-edge participates in a crossing in $\mathcal{G}$.
\end{obs}
\begin{proof}
	Suppose, for the sake of contradiction, that $\mathcal{G}$ contains a crossing involving an \HB-edge.
	Any such crossing contributes at least the weight of one \HB-edge to the crossing number, that is, at least $\omega^{8}$.
	Since~$\omega^{8} > k$, this implies that $\crn(\mathcal{G}) > k$, contradicting the assumption that $\crn(\mathcal{G}) \leq k$.
\end{proof}

\begin{obs}\label{obs:subgraph}
	Let $H$ be a subgraph of $G$. If $(H,k)$ is a no-instance of \textsc{Crossing Number}, then $(G,k)$ is also a no-instance of \textsc{Crossing Number}.
\end{obs}
\begin{proof}
	Assume the contrary, that $(G,k)$ is a yes-instance of \textsc{Crossing Number}.
	Then there exists a drawing~$\mathcal{G}$ of~$G$ such that $\crn(\mathcal{G})\leq k$.
	By deleting from $\mathcal{G}$ all vertices and edges of $G \setminus H$, we obtain a drawing~$\mathcal{H}$ of~$H$.
	Since deleting vertices or edges cannot increase the number of crossings, we have $\crn(\mathcal{H}) \leq \crn(\mathcal{G}) \leq k$.
	This yields a solution for $(H,k)$, contradicting the assumption that $(H,k)$ is a no-instance.
\end{proof}

Based on~\cref{obs:subgraph}, we may restrict our attention to suitable subgraphs of $G$ when deriving necessary conditions for drawings with at most $k$ crossings.
In particular, as defined in \cref{sub:auxill}, let $F$ denote the frame and let $G'$ denote the graph consisting of the frame together with $n$ variable $\Var$-gadgets (see \cref{fig:auxill}).
We therefore focus on properties of drawings that must hold already for $G'$, independently of how the clauses are encoded in the full graph~$G$.

The next two claims are proved using the same underlying argument: any forbidden crossing between the considered edges would incur a cost exceeding the prescribed bound $k$, and hence cannot occur in any drawing $\mathcal{G}$ with~$\crn(\mathcal{G}) \leq k$.

\begin{claim}\label{cl:varcrossing}
	If there exists a drawing $\mathcal{G}'$ of $G'$ such that $\crn(\calG')\leq k$, then the only possible crossings in $\mathcal{G}'$ are crossings between an \mbox{\Rr-} or \Bb-edge and an \Rrh- or \Bbh-edge, or between two \mbox{\Rrh-} or \Bbh-edges.
	Consequently, no two vertical paths of the same or different $\Var$-gadgets may cross each other or self-cross.
\end{claim}
\begin{proof}
	Suppose that $\mathcal{G}'$ contains a crossing in which both edges have weight greater than $\omega^{3}$.
	Since all such edges have weight at least $\omega^{4}$, the contribution of this crossing to the crossing number is at least $\omega^{4}\cdot \omega^{4} = \omega^{8}$.
	As $\omega^{8} > k$, this contradicts the assumption that $\crn(\mathcal{G}') \leq k$.
	
	Hence, any crossing in $\mathcal{G}'$ must involve at least one edge of weight $\omega^{3}$, that is, an \Rrh- or a \Bbh-edge.
	However, \LB-edges cannot participate in any crossing either: even a crossing between an \LB-edge (of weight at least $\omega^{6}$) and an edge of weight $\omega^{3}$ contributes at least $\omega^{9}$, which again exceeds $k$.
	
	Therefore, the only crossings that may occur in $\mathcal{G}'$ without violating the bound $\crn(\mathcal{G}') \leq k$ are crossings between an \Rr- or \Bb-edge and an \Rrh- or \Bbh-edge, or between two \Rrh- or \Bbh-edges.
	Any crossing between vertical paths of $\Var$-gadgets would yield a crossing forbidden by this condition.
\end{proof}

\begin{claim}\label{cl:face}
	If a drawing $\calG$ of $G$ is a solution of the instance $(G, k)$,
	then all vertices $u, v\in V(G'\setminus F)$ and all \Gg-edges of~$G$ are drawn in the same face of the frame $F$ in $\mathcal{G}$, and this face is uniquely determined.	
\end{claim}
\begin{proof}
	By \cref{obs:HB}, the subdrawing of the frame $F$ in $\calG$ is uncrossed.
	Furthermore, observe that, up to a choice of the outer face, there is a unique plane embedding of $F$ such that all vertices of the set~$\big\{r_{j, R}^0, r_{j, L}^{n+1}\big\}_{j\in[2, h+2]}$ are incident to the same face, which we denote by~$\phi$.
	Likewise, there is a unique plane embedding of $F$ such that both vertices of the set~$\{c_{j,L},c_{j,R}\}$, for any ${j\in[\ell]}$, are incident to the same face; moreover, it is again the same embedding and the same face~$\phi$.
	
	Let $\mathcal{F}$ denote the embedding of the frame $F$ induced by the drawing $\mathcal{G}$.
	Suppose, for the sake of contradiction, that there exists a vertex $u \in V(G' \setminus F)$ that lies in a face of $\mathcal{F}$ distinct from~$\phi$.
	Since the graph $G' \setminus F$ is connected, $u$ is connected by a path to every vertex in $\big\{r_{j, R}^0, r_{j, L}^{n+1}\big\}_{j\in[2, h+2]}$, and some of these paths hence must cross an edge of the frame.
	However, all edges of the frame are \HB-edges, and by \cref{obs:HB}, no \HB-edge can participate in a crossing in any drawing with crossing number at most $k$.
	This yields a contradiction.
	
	For every \Gg-edge $c_{j,L}c_{j,R}$ where $j\in[\ell]$, the same argument applies: unless some \HB-edge of the drawing~$\mathcal{F}$ is crossed by $c_{j,L}c_{j,R}$ (which contradicts \cref{obs:HB}), the edge $c_{j,L}c_{j,R}$ must be drawn in the unique face~$\phi$ of $\mathcal{F}$.
	%
\end{proof}

By \cref{cl:face}, we may, without loss of generality, fix a drawing of the frame $F$ with the~$\Var$-gadgets as shown in~\cref{fig:frame}, that is, with all $\Var$-gadgets drawn in the same face of the subdrawing of~$F$.
In particular, this fixes the left and right sides of the frame and thereby induces a linear ordering of the $\Var$-gadgets from left to right, following the increasing order of indices $i \in [n]$.
Moreover, this choice, by \cref{cl:varcrossing}, almost completely determines the placement of the $\Var$-gadgets inside the frame, up to possible crossings involving one or two \Rrh- and \Bbh-edges:

\begin{claim}\label{cl:varplacement}
	If there exists a drawing $\mathcal{G}'$ of $G'$ such that $\crn(\calG')\leq k$, then the following is true in $\mathcal{G}'$ for all~$i\in [n]$:
	\begin{enumerate}[a)]
		\item the \texttt{R-left} path of the gadget $\Var^i$ is drawn to the left of both \LB-paths of $\Var^i$, while the \texttt{R-right} path of $\Var^i$ is drawn to the right of them,
		\item either the \texttt{B-pos} path of the gadget $\Var^i$ is drawn to the left of its \texttt{R-left} path and the \texttt{B-neg} path of $\Var^i$ is to the right of its \texttt{R-right} path,
		or the \texttt{B-neg} path of the gadget~$\Var^i$ is drawn to the left of its \texttt{R-left} path and the \texttt{B-pos} path of $\Var^i$ is to the right of its \texttt{R-right} path, and
		\item the whole gadget $\Var^{i-1}$ (if~$i>1$) is drawn to the left of each of the paths \texttt{B-pos}, \texttt{B-neg} of $\Var^i$, and the whole gadget $\Var^{i+1}$ (if~$i<n$) is drawn to the right of them.
	\end{enumerate}
\end{claim}
\begin{proof}a)
	Assume the contrary, for some $i\in [n]$. By \cref{cl:varcrossing}, no two vertical paths of~$\Var^i$ cross, this means that the \texttt{R-left} path is drawn to the right of both \LB-paths of~$\Var^i$.
	By the construction (see \cref{sec:frame} and \cref{fig:auxill} for an illustration), the \texttt{R-left} path has \mbox{\Rrh-edges} (the "stairs") connecting it to the \texttt{R-right} path of~$\Var^{i-1}$ (or, if~$i=1$, to a subdivision of the frame's side $u^{BL} u^{TL}$). 
	Since the \LB-paths of $\Var^i$ connect the top and bottom sides of the frame and must remain inside the preselected face of the frame (by \cref{cl:face}), the~\texttt{R-left} path would be forced to cross an \HB-edge or an \LB-edge of $\Var^i$, which violates \cref{cl:varcrossing}.
	The same reasoning applies symmetrically for the \texttt{R-right} path of $\Var^i$.
	
	b) Again, by \cref{cl:varcrossing}, no two vertical paths of $\Var^i$ cross.
	So, for the sake of contradiction, we may assume that both \texttt{B-pos} and \texttt{B-neg} paths of $\Var^i$ are to the right of both \LB-paths of $\Var^i$, or symmetrically, both \texttt{B-pos} and \texttt{B-neg} paths are to the left.
	However, at least one internal vertex on the~\texttt{B-pos} path is adjacent via a \Bbh-edge (horizontal) to an internal vertex on the \texttt{LB-pos} path of $\Var^i$,
	and at least one internal vertex on the~\texttt{B-neg} path is adjacent via a \Bbh-edge to an internal vertex on the~\texttt{LB-neg} path.
	Hence, depending on their mutual order, one of the \LB-paths of $\Var^i$ must be crossed by a \Bbh-edge, which violates \cref{cl:varcrossing}.
	
	c) This now follows directly from \cref{cl:varcrossing}.
	%
\end{proof}


Next, again using \cref{cl:varcrossing}, we account for crossings enforced between the \Gg-edges of~$G$ and the subdrawing of~$G'$, which exist regardless of the \textsc{Satisfiability} instance considered in our reduction.

\begin{claim}\label{cl:green}
	If a drawing $\calG$ of $G$ is a solution of the instance $(G, k)$,
	then crossings between the \Gg-edges of~$G$ (the clause edges) and edges of the subgraph $G'$ in $\calG$ contribute in total at least $2n\ell\omega^6+4n\ell\omega^4$ to the count~$\crg(\calG)$.
	In particular, every \Gg-edge in $\calG$, for each $i\in[n]$, crosses the $\Var^i$ gadget in (at least) two \LB-edges, two \Bb-edges and two \Rr-edges.
\end{claim}
\begin{proof}
	This follows easily from \cref{cl:varplacement} and \cref{obs:HB}.
	There are altogether $2n$ \LB-paths of all variable $\Var$-gadgets and $4n$ such \Bb- and \Rr-paths, all pairwise edge-disjoint. By the Jordan curve theorem, since the \HB-edges remain uncrossed, each of these paths have to be crossed by each of the $\ell$ \Gg-edges of~$G$. 
	By \cref{tab:colors}, these crossings contribute in total at least $2n\ell\omega^6+4n\ell\omega^4$.
\end{proof}

\begin{corollary}\label{cor:green}
	If a drawing $\calG$ of $G$ is a solution of the instance $(G, k)$, then (cf.~\cref{cl:varcrossing}) the total contribution of all crossings between \Rr-, \Bb-, \Rrh- and \Bbh-edges in $\calG$ to the count~$\crg(\calG)$ is at most
	\begin{equation*}
	2n\!\left(\!(2h+1)\omega^7+\sum\limits_{j=2}^{h+1}j(j+1)\omega^4+\sum\limits_{j=1}^{h+1}j(j+2)\omega^4\!\right)+\mathcal{O}_{n,\ell}(\omega^2).
	\end{equation*}
\end{corollary}
\begin{proof}
	This follows by comparing \eqref{eq:kvalue} to \cref{cl:green}.
\end{proof}

Finally, we argue that the vertical placement of the stairs connecting the \Rr-paths of neighboring~$\Var$-gadgets is fixed as well, completing the essential constraints imposed on the drawing.
Recall that, during the construction, we have introduced a small \emph{adjustment} weight of order $\omega^1$ on the vertical \Rr- and \Bb-edges (cf.~\cref{tab:colors}).
This careful selection of adjustment weights enforces a unique alternation pattern of crossings, so that any optimal drawing of $G$ must realize the crossings exactly as illustrated in \cref{fig:frame} and the subsequent figures.

\begin{lemma}\label{lemma:w7}
	If a drawing $\calG$ of $G$ is a solution of the instance $(G, k)$, then each horizontal \Bbh-edge \mbox{(\Rrh-edge)} crosses exactly one vertical \Rr-edge (\Bb-edge). 
	Moreover, the first and the last edge of every vertical \Rr-path are not crossed by \mbox{\Bbh-edges}.
	The total contribution of all such crossings to the count~$\crg(\calG)$ is exactly
	\begin{equation}\label{eq:crstairs}
	2n\!\left(\!(2h+1)\omega^7+\sum\limits_{j=2}^{h+1}j(j+1)\omega^4+\sum\limits_{j=1}^{h+1}j(j+2)\omega^4\!\right).
	\end{equation}
\end{lemma}
\begin{proof}
	Let $\calG'$ denote the subdrawing of the graph $G'$ in~$\calG$.
	We first observe that the claimed~\eqref{eq:crstairs} total weight of all crossings between \Rrh- (\Bbh-) edges (the stairs) and the corresponding \mbox{\Bb-} (\Rr-) edges is achieved in $\calG'$ when these crossings  alternate exactly along the vertical paths and avoid the first and the last edge of every vertical \Rr-path---informally, when the picture is exactly as illustrated in \cref{fig:frame}.
	Our goal is to show that any deviation from this alternating pattern increases the total crossing weight, due to the assigned adjustment weights of the vertical \Rr- and \Bb-edges by at least $\omega^4>\mathcal{O}_{n,\ell}(\omega^2)$, thereby violating \cref{cor:green}.
	\begin{figure}[h]
		\input{base}
		\caption{Illustration for the proof of~\cref{lemma:w7}}
		\label{fig:base}
	\end{figure}
	
	Recall that, for all $x\in \{L, R\}$, $i\in [n]$:
	\begin{itemize}
		\item for $j\in [h+2]$, the edge $r_{j, x}^{i}r_{j+1, x}^{i}$ has weight $\omega^4+j(j+1)\omega$, and denote this value by~$g_j=\omega^4+j(j+1)\omega$;
		\item for $j\in [h+1]$, the edge $b_{j, x}^{i}b_{j+1, x}^{i}$ has weight $\omega^4+j(j+2)\omega$, and denote this value by~$s_j=\omega^4+j(j+2)\omega$.
	\end{itemize}
	Now we observe that, by \cref{cl:varplacement}, every \Bbh-edge crosses the neighboring \Rr-path of the same $\Var$-gadget, and every $\Rrh$-edge likewise crosses the neighboring \Bb-path, and so
	the total contribution of the crossings between \Rrh-edges and \Bb-edges, and between \Bbh-edges and \Rr-edges, is at least $2n\big((h+1)\omega^3\omega^4+h\omega^3\omega^4\big)=2n(2h+1)\omega^7$.
	Comparing this lower bound to \cref{cor:green}, we conclude that every \Bbh-edge (\Rrh-edge) in $\calG'$ crosses only the corresponding one \Rr-path (\Bb-path), and these are all crossings in $\calG'$ allowed by \cref{cl:green}.
	We rely on this obervation and on the ordering of vertical paths described by \cref{cl:varplacement} in the rest of the proof.
	
	We aim to show that each of the $2n$ pairs of neighboring \Rr- and \Bb-paths together with their corresponding stairs, contributes at least 
	$A(h):=(2h+1)\omega^7+\sum\limits_{j=2}^{h+1}j(j+1)\omega^4+\sum\limits_{j=1}^{h+1}j(j+2)\omega^4$ to the total crossing number. This then establishes the lower bound \eqref{eq:crstairs} and specifies the conditions under which equality can hold.
	
	We proceed by induction on the height $h$.
	For $h=1$, there are exactly three possible relative placements of crossings between the \Rr- and \Bb-edges. 
	As it is shown in~\cref{fig:base}, only the alternating placement achieves the minimum possible contribution, namely $3\omega^7+17\omega^4$, which coincides with the claimed formula~$A(1)$.
	
	Assume that for some integer $h\geq1$, the crossings between a neighboring pair of \Rr- and \Bb-paths, together with their stairs, contribute at least $A(h)$ to the total crossing number.
	
	We now increase the height of our gadgets to $h+1$. 
	This in each staircase of our pair of \Rr- and \Bb-paths introduces, up to symmetry of the $\Var$-gadgets, one additional horizontal \Bbh-edge $f_1=b_{h+2, P}^i v_{h+2, P}^i$ and one additional horizontal \mbox{\Rrh-edge} $f_2=r_{h+3, R}^{i-1}r_{h+3, L}^{i}$.
	The vertical \Rr- and \Bb-paths are accordingly prolonged each by one new edge of weight~$g_{h+3}$ and~$s_{h+2}$, respectively, which are heavier than other edges in these paths.
	
	By the induction hypothesis, the contribution of our pair of \Rr- and \Bb-paths with their stairs, but not counting $f_1$ and~$f_2$, to the crossing number is at least~$A(h)$.
	We now have two possibilities (cases); (i) the edge $f_2$ crosses the vertical \Bb-path above the vertex $b_{h+2,P}^i$, or (ii) the edge $f_1$ crosses the vertical \Rr-path above the vertex $r_{h+3,L}^{i}$.
	
	In Case (i), the \Rrh-edge $f_2$ crosses the \Bb-edge $b_{h+2,P}^i b_{h+3,P}^i$ of weight $s_{h+2}$, which contributes $\omega^3s_{h+2}$ crossings above~$A(h)$.
	Furthermore, the \Bbh-edge $f_1$ crosses an \Rr-edge $r_{j, x}^{i}r_{j+1, x}^{i}$ of the neighboring \Rr-path of weight~$g_j$ for some $j\in [h+2]$, which contributes $\omega^3g_j$ crossings.
	However, in such case we also get that~$(h+2-j)$ of the \Rrh-edges other than~$f_2$ cross the neighboring \Bb-path in the \Bb-edge~$b_{h+2,P}^i b_{h+3,P}^i$ of weight~$s_{h+2}$, instead of crossing it in \Bb-edges of weight at most~$s_{h+1}$ as accounted for in the expression~$A(h)$ by induction.
	The total contribution to the crossing number for height $h+1$ thus in this case is at least
	\begin{eqnarray*}
		A(h)\!\!\!&+&\!\!\!\omega^3s_{h+2}+\omega^3g_j+(h+2-j)(s_{h+2}-s_{h+1})
		\\	 &=& A(h)+2\omega^7+(h+2)(h+4)\omega^4+j(j+1)\omega^4+(h+2-j)(2h+5)\omega^4
		\\	 &=& A(h)+2\omega^7+ (3h^2+15h-2jh+j^2-4j+18)\omega^4
		\\	 &=& A(h)+2\omega^7+ \big(2h^2+11h+18+ (h+4-j)(h-j)\big)\omega^4
		.\end{eqnarray*}
	Since $(h+4-j)>0$ for any $j\in [h+2]$, the last expression can be estimated from below by
	\begin{eqnarray*}
		A(h)\!\!\!&+&\!\!\!2\omega^7+ \big(2h^2+11h+18+ (h+4-j)(h-j)\big)\omega^4
		\\	 &\geq& A(h)+2\omega^7+ \big(2h^2+11h+18+ (h+4-(h+2))(h-(h+2))\big)\omega^4
		\\	 &=& A(h)+2\omega^7+ (2h^2+11h+14)\omega^4
		\\	 &=& A(h)+2\omega^7+ (h+2)(h+3)\omega^4+(h+2)(h+4)\omega^4 =A(h+1)
		,\end{eqnarray*}
	where equality holds if and only if~$j=h+2$, that is, if and only if the crossings again precisely alternate along the vertical paths.
	In all other subcases, the contribution is at least~$A(h+1)+\omega^4$ which violates \cref{cor:green}.
	
	In Case (ii), we analogously estimate the total contribution to the crossing number. In this case, for some~$j\in[h+1]$, since the edge $f_2$ this time crosses the \Bb-path below the vertex~$b_{h+2,P}^i$,  the total contribution is at least
	\begin{eqnarray*}
		A(h)\!\!\!&+&\!\!\!\omega^3g_{h+3}+\omega^3s_j+(h+1-j)(g_{h+3}-g_{h+2})
		\\	 &=& A(h)+2\omega^7+(h+3)(h+4)\omega^4+j(j+2)\omega^4+(h+1-j)(2h+6)\omega^4
		\\	 &=& A(h)+2\omega^7+ \big(2h^2+11h+18+ (h+4-j)(h-j)\big)\omega^4
		\\	 &\geq& A(h)+2\omega^7+ \big(2h^2+11h+18+ (h+4-(h+1))(h-(h+1))\big)\omega^4
		\\	 &=& A(h)+2\omega^7+ (2h^2+11h+15)\omega^4
		\\	 &=& A(h)+2\omega^7+ (h+2)(h+3)\omega^4+(h+2)(h+4)\omega^4 +\omega^4 =A(h+1)+\omega^4
		.\end{eqnarray*}
	We can see that no subcase of Case (ii) is admissible by \cref{cor:green}.
	%
	%
	%
\end{proof}

Now, let us examine the low-order terms, namely $n\ell\omega^2+ (\omega^2-1)$, in the expression \eqref{eq:kvalue}, which, together with the contribution captured by \cref{cl:green}, come from the crossings of the \Gg-edges with the loaded $\Var_+$-gadgets in~$G$ (cf.~\cref{sub:fullred} for the full description).

\begin{claim}\label{cl:green_loaded}
	If a drawing $\calG$ of $G$ is a solution of the instance $(G, k)$, then the crossings of~$\calG$ incident to the \Gg-edges altogether contribute at least $2n\ell\omega^6+4n\ell\omega^4+n\ell\omega^2$ to the crossing number.
	In particular, every \Gg-edge in $\calG$, for each $i\in[n]$, crosses the loaded variable gadget~$\Var_+^i$ also in an edge not belonging to~$\Var^i$.
\end{claim}
\begin{proof}
	A lower bound of $2n\ell\omega^6+4n\ell\omega^4$ on the contribution of crossings between the \Gg-edges and the $\Var$-gadgets is shown in \cref{cl:green}.
	Now, we observe that the construction of $\Var_+^i$ in \cref{sub:fullred} introduces a path which is edge-disjoint from $\Var^i$ and connects \HB-edges at the top and the bottom parts of the frame~$F$.
	Again, by the Jordan curve theorem, this path has to be crossed by each \Gg-edge in $\calG$.
	Since every edge of this path is of weight at least that of a \Cc-edge, i.e.~$\omega^2$, we get the sought additional crossing contribution of at least $n\ell\omega^2$.
\end{proof}

\begin{corollary} \label{cor:ccedges}
	If a drawing $\calG$ of $G$ is a solution of the instance $(G, k)$, then
	\begin{itemize}
		\item[--] no \LB- or \Cc-edge of any $\Var_+$-gadget is crossed by an edge other than some \Gg-edge in $\calG$, and
		\item[--] each \Gg-edge in $\calG$ crosses each $\Var_+$-gadget in exactly two \LB-edges, two \Bb-edges, two \Rr-edges and one \Cc-edge.
	\end{itemize}
\end{corollary} 
\begin{proof}
	By combining \cref{lemma:w7} and \cref{cl:green_loaded}, we get a lower bound on the number of crossings of $\cal G$ which differs from the crossing budget $k$ in \eqref{eq:kvalue} by only $(\omega^2-1)$.
	Since every edge of every $\Var_+$-gadget is of weight at least $\omega^2$, we cannot have more crossings in any $\Var_+$-gadget than the bare minimum accounted for in \cref{lemma:w7} and \cref{cl:green_loaded}, which is the sought conclusion.
\end{proof}

We continue with analyzing the low-order terms of \eqref{eq:kvalue}, with the goal to relate a solution~$\calG$ of the instance~$(G, k)$ to a satisfying assignment of the given \textsc{Satisfiability} instance.
In a nutshell, we aim to (formulate and) prove that every \Gg-edge of $\calG$ has to be routed in $\calG$ within its corresponding grey-shaded area as illustrated in \cref{fig:example-noflip} and subsequent pictures of the reduction.

For the coming arguments, we need the following special notation for the vertical \LB- and \Bb-paths of the~$\Var$-gadgets with respect to a fixed drawing $\calG$ of $G$ which is a solution of the instance $(G, k)$.
If the~\texttt{LB-pos} path $Q$ of $\Var^i$ is drawn to the right of its \texttt{LB-neg} path, then let~$v_{j,R}^i$ denote the vertex~$v_{j,P}^i$ of~$Q$ and~$v_{j,L}^i$ denote the vertex~$v_{j,N}^i$ of~$Q$.
Otherwise, let~$v_{j,R}^i$ denote the vertex~$v_{j,N}^i$ and~$v_{j,L}^i$ denote the vertex~$v_{j,P}^i$.
The notation $b_{j,R}^i$ and $b_{j,L}^i$ is introduced analogously for the vertices of the two \Bb-paths of $\Var^i$.

\begin{claim}\label{cl:onlycell}
	Let $\calG$ of $G$ be a solution of the instance $(G, k)$.
	If a \Gg-edge in $\calG$ crosses the vertical \LB-edge $e=v_{j,L}^i v_{j+1,L}^i$ of $\Var^i$, for $i\in[n]$ and $j\in[4\ell+n-1]$, then $e$ has both ends in a cell of the loaded variable gadget $\Var_+^i$.
\end{claim}
\begin{proof}
	Observe from \cref{cor:ccedges} that the subdrawing of $\calG$ induced by the \LB- and \Cc-edges of each $\Var_+$-gadget is a unique plane embedding.
	So, by the construction in \cref{sub:fullred}, the considered \Gg-edge $f$, after crossing $e$, has to cross one of the \LB- or \Cc-edges of $E(\Var_+^i)\setminus E(\Var^i)$.
	However, again by \cref{cor:ccedges}, this edge crossed by $f$ must be a \Cc-edge, and the conclusion follows.
\end{proof}

\begin{claim}\label{cl:notup}
	Let $\calG$ of $G$ be a solution of the instance $(G, k)$, let $f$ be a \Gg-edge in $\calG$, and $i\in[n-1]$, $j\in[4\ell+n-3]$.
	If $f$ crosses the vertical \LB-edge $v_{j,R}^i v_{j+1,R}^i$ in $\Var^i$, then $f$ cannot cross any of the \LB-edges $v_{j',L}^{i+1} v_{j'+1,L}^{i+1}$ in $\Var^{i+1}$ for $j'\geq j+2$.
\end{claim}
\begin{figure}[h]
	\vspace*{-2ex}	\centering
\begin{tikzpicture}[xscale=1.3,yscale=1.3]\footnotesize
\small
\tikzstyle{every path}=[draw]
\tikzstyle{every node}=[color=red\blackpart]

\tikzstyle{every node}=[draw, color=blue\blackpart, shape=circle, inner sep=1.1pt, fill=blue\blackpart]
\def\xx{2}
\draw[semithick,blue\blackpart] (\xx,-0.7)--(\xx,3.8);
\draw[uweight,blue\blackpart] (\xx,-0.7)--(\xx,3.8);
\foreach \yy in {0,...,3}  \draw[blue\blackpart] (\xx,\yy)-- ++(-2,0);
\node[label=right:$b^i_{j+3,R}$] at (\xx,3) {};
\node[label=right:$b^i_{j+2,R}$] at (\xx,2) {};
\node[label=right:$b^i_{j+1,R}$] at (\xx,1) {};
\node[label=right:$b^i_{j,R}$] at (\xx,0) {};
\def\xx{4}
\draw[semithick,blue\blackpart] (\xx,-0.7)--(\xx,3.8);
\draw[uweight,blue\blackpart] (\xx,-0.7)--(\xx,3.8);
\foreach \yy in {0,...,3}  \draw[blue\blackpart] (\xx,\yy)-- ++(2,0);
\node[label=left:$b^{i+1}_{j+3,L}$] at (\xx,3) {};
\node[label=left:$b^{i+1}_{j+2,L}$] at (\xx,2) {};
\node[label=left:$b^{i+1}_{j+1,L}$] at (\xx,1) {};
\node[label=left:$b^{i+1}_{j,L}$] at (\xx,0) {};

\tikzstyle{every node}=[draw, color=red\blackpart, shape=circle, inner sep=1.1pt, fill=red\blackpart]
\def\xx{1}
\draw[semithick,red\blackpart] (\xx,-0.7)--(\xx,3.8);
\draw[uweight,red\blackpart] (\xx,-0.7)--(\xx,3.8);
\foreach \yy in {-1,...,3}  \draw[red\blackpart] (\xx,\yy+0.5)-- ++(4,0);
\node[label=left:$r^i_{j+4,R}$] at (\xx,3.5) {};
\node[label=left:$r^i_{j+3,R}$] at (\xx,2.5) {};
\node[label=left:$r^i_{j+2,R}$] at (\xx,1.5) {};
\node[label=left:$r^i_{j+1,R}$] at (\xx,0.5) {};
\node[label=left:$r^i_{j,R}$] at (\xx,-0.5) {};
\def\xx{5}
\draw[semithick,red\blackpart] (\xx,-0.7)--(\xx,3.8);
\draw[uweight,red\blackpart] (\xx,-0.7)--(\xx,3.8);
\node[label=right:$r^{i+1}_{j+4,L}$] at (\xx,3.5) {};
\node[label=right:$r^{i+1}_{j+3,L}$] at (\xx,2.5) {};
\node[label=right:$r^{i+1}_{j+2,L}$] at (\xx,1.5) {};
\node[label=right:$r^{i+1}_{j+1,L}$] at (\xx,0.5) {};
\node[label=right:$r^{i+1}_{j,L}$] at (\xx,-0.5) {};

\tikzstyle{every node}=[draw, color=black, shape=circle, inner sep=1.1pt, fill=black]
\def\xx{0}
\draw[thick,black] (\xx,-0.7)--(\xx,3.8);
\node[label=left:$v^i_{j+3,R}$] at (\xx,3) {};
\node[label=left:$v^i_{j+2,R}$] at (\xx,2) {};
\node[label=left:$v^i_{j+1,R}$] at (\xx,1) {};
\node[label=left:$v^i_{j,R}$] at (\xx,0) {};
\def\xx{6}
\draw[thick,black] (\xx,-0.7)--(\xx,3.8);
\node[label=right:$v^{i+1}_{j+3,L}$] at (\xx,3) {};
\node[label=right:$v^{i+1}_{j+2,L}$] at (\xx,2) {};
\node[label=right:$v^{i+1}_{j+1,L}$] at (\xx,1) {};
\node[label=right:$v^{i+1}_{j,L}$] at (\xx,0) {};

\normalsize
\tikzstyle{every path}=[draw,semithick, color=green\gblackpart, rounded corners=4pt]
\draw (-1,0.71) node[draw=none,fill=none, label=left:$f\!\!$] {} --++(3.75,0) --++(0.5,0.6) --++(3.5,0);
\draw[dashed] (2.6,0.7) --++(4,0);
\draw[dashed] (-0.1,0.7) --++(0.5,-0.5) --++(2.2,0) --++(0.5,-0.5) --++(3.5,0);

\node[draw=none,fill=none] at (-0.8,3.8) {$\Var^i$};
\node[draw=none,fill=none] at (6.8,3.8) {$\Var^{i+1}$};

\end{tikzpicture}
	\caption{Illustration for the proof of~\cref{cl:notup}; possible routings of the \Gg-edge~$f$.}
	\label{fig:notup}
\end{figure}
\begin{proof}
	Since $f$ cannot cross the horizontal \Bbh-edges $v_{j,R}^i b_{j,R}^i$ and $v_{j+1,R}^i b_{j+1,R}^i$ by \cref{cor:ccedges}, $f$ has to cross the vertical \Bb-edge $b_{j,R}^i b_{j+1,R}^i$.
	By \cref{lemma:w7}, the vertical \Bb-edges $b_{j+1,R}^i b_{j+2,R}^i$ and $b_{j,L}^{i+1} b_{j+2,L}^{i+1}$ are both crossed by the same horizontal \Rrh-edge $r_{j+2,R}^i r_{j+2,L}^{i+1}$.
	Consequently, the edge $f$ cannot cross any of the \Bb-edges $b_{j',L}^{i+1} b_{j'+1,L}^{i+1}$ for $j'\geq j+2$ (see details in \cref{fig:notup}).
	And since, again, $f$ cannot cross any of the horizontal \mbox{\Bbh-edges} $b_{j',L}^{i+1} v_{j',L}^{i+1}$ by \cref{cor:ccedges}, the conclusion follows.
\end{proof}

\begin{corollary} \label{cor:cellsarea}
	If a drawing $\calG$ of $G$ is a solution of the instance $(G, k)$ and $f\in E(G)$ is the \Gg-edge of the clause $C_j\in\mathcal{C}$ for $j\in[\ell]$, then
	\begin{enumerate}[a)]
		\item $f$ crosses in $\calG$ each loaded variable gadget $\Var_+^i$ in its cell of $C_j$, and
		\item specifically, there is $i\in[n]$ such that $f$ crosses $\Var^i$ in the edges $v^i_{4j+i-3,L} v^i_{4j+i-2,L}$ and $v^i_{4j+i-2,R} v^i_{4j+i-1,R}$
		(informally, $f$ ``jumps up'' through the cell of $C_j$ in $\Var_+^i$, see \cref{fig:example-assignments}).
	\end{enumerate}
\end{corollary} 
\begin{proof}a)
	By the construction of $G$, the edge $f$ starts in the vertex $c_{j, L}$which is the vertex with neighbors~$r^0_{4j-2, R}$ and~$r^0_{4j-1, R}$ on the left side of the frame.
	Then, by \cref{cor:ccedges} and \cref{cl:onlycell}, $f$ can cross the left side of the gadget $\Var^1$ only in the edge $v^1_{4j-2,L} v^1_{4j-1,L}$, and hence in the cell of $C_j$ in $\Var_+^1$.
	By a symmetric argument, since $f$ ends in a vertex with neighbors $r^{n+1}_{4j+n-1, L}$ and $r^{n+1}_{4j+n, L}$ on the right side of the frame,
	$f$ can cross the right side of the gadget $\Var^n$ only in the edge $v^n_{4j+n-2,R} v^1_{4j+n-1,R}\,$, and hence again in the cell of $C_j$ in~$\Var_+^n$.
	
	By \cref{cl:notup}, if $f$ crosses a cell of $C_{j'}$ in $\Var_+^i$, then $f$ can only cross a cell of $C{j''}$ in~$\Var_+^{i+1}$ such that~$j''\leq j'$.
	Consequently, a cell in which $f$ crosses each gadget $\Var_+^i$ has to always be the cell of~$C_j$.
	
	b) Assume the contrary; hence, if $f$ crosses $\Var^i$ in the edge $v^i_{4j+i-3,L} v^i_{4j+i-2,L}$, then $f$ crosses $\Var^i$ also in the edge $v^i_{4j+i-3,R} v^i_{4j+i-2,R}$.
	We show by induction on $i$ that $f$ crosses every $\Var^i$ in the edges $v^1_{4j+i-3,L} v^1_{4j+i-2,L}$ and $v^1_{4j+i-3,R} v^1_{4j+i-2,R}$.
	This is true for $i=1$ by the argument of a), and the induction step follows from \cref{cl:notup} and \cref{cl:onlycell}.
	Finally, the edge $f$ crosses $\Var^n$ in the edge $v^1_{4j+n-3,R} v^1_{4j+n-2,R}\,$, but this contradicts the arguments in a).
\end{proof}

\subsection{Correctness}\label{subsec:correctness}

Having defined our reduction from \textsc{Satisfiability} to \textsc{Crossing Number} and shown its technical properties in the previous subsections,
we are now in a position to finish the proof of the first part \cref{thm:mainred}.

\begin{lemma}\label{lem:redforward}
	If $\mathcal{I}=(\mathcal{C}, \mathcal{V})$ is a positive instance of \textsc{Satisfiability}, then the graph $G$ of the constructed instance $(G,k)$ admits a drawing $\calG$ in the plane such that $\crn(\calG)\leq k$.
\end{lemma}

\begin{proof}
	Let $\tau:\mathcal{V}\to\{\false,\true\}$ be a satisfying truth assignment of the instance $\mathcal{I}$.
	Let a graph $G'$ be the frame with $n$ variable $\Var$-gadgets, which is a subgraph of~$G$.
	We consider a drawing $\calG'$ of $G'$ as specified by \cref{fig:auxill}, additionally satisfying the following for every $i\in[n]$;
	\begin{itemize}
		\item[--] if $\tau(x_i) = \true$, then the subdrawing of $\Var^i$ in $\calG'$ is flipped such that the \texttt{LB-pos} path is to the left of \texttt{LB-neg};
		\item[--] if $\tau(x_i) = \false$, then the subdrawing of $\Var^i$ in $\calG'$ is flipped such that the \texttt{LB-neg} path is to the left of \texttt{LB-pos}.
	\end{itemize}
	Then we extend the drawing $\calG'$ to a drawing $\calG'_+$ of $G'$ with all loaded variable gadgets, as defined in \cref{sub:fullred} and specifically in \cref{fig:cells}.
	The drawing $\calG'_+$ looks as shown in \cref{fig:example-noflip} (so far without the \Gg-edges).
	
	By \cref{lemma:w7}, we have $\crn(\calG'_+)=2n(2h+1)\omega^7+2n\sum\limits_{j=2}^{h+1}j(j+1)\omega^4+2n\sum\limits_{j=1}^{h+1}j(j+2)\omega^4$.
	
	Consider now a clause $C_j\in \mathcal{C}$. Since we have a positive instance of \textsc{Satisfiability}, $C_j$ is satisfied by at least one of its literals, say by one containing the variable~$x_m\in\mathcal{V}$.
	We draw the \Gg-edge $f_j$ of $C_j$, from its end $c_{j,L}$ on the left side of the frame to its end $c_{j,R}$ on the right side as follows.
	\begin{itemize}
		\item For all $i\in[m-1]$, we draw $f_j$ crossing the gadget $\Var_+^i$ in the ``lower half'' of the cell of~$C_j$; precisely, the edge~$f_j$ crossing the edges $v^i_{4j+i-3,L} v^i_{4j+i-2,L}$ and~$v^i_{4j+i-3,R} v^i_{4j+i-2,R}$ of $\Var^i$ and the (single) \Cc-edge of the cell of $C_j$ between them (cf.~\cref{fig:cells}).
		\item We draw $f_j$ crossing the gadget $\Var_+^m$ in the edges $v^m_{4j+m-3,L} v^m_{4j+m-2,L}$ and $v^m_{4j+m-2,R} v^m_{4j+m-1,R}$ (``jumping from the lower to the upper half'' of the cell of $C_j$ there).
		If $x_m\in C_j$ (and $x_m=\true$ to satisfy~$C_j$), then this routing of $f_j$ is possible with crossing only one \Cc-edge by \cref{fig:cells}(a).
		If $\overline{x_m}\in C_j$ (and $x_m=\false$ to satisfy~$C_j$), then this routing of $f_j$ is again possible with crossing only one \Cc-edge by \cref{fig:cells}(b) which is vertically flipped in~$\calG'_+$.
		\item For all $i\in[m+1,n]$, we draw $f_j$ crossing the gadget $\Var_+^i$ in the ``upper half'' of the cell of~$C_j$; precisely, the edge~$f_j$ crossing the edges $v^i_{4j+i-2,L} v^i_{4j+i-1,L}$ and $v^i_{4j+i-2,R} v^i_{4j+i-1,R}$ of $\Var^i$ and the (single) \Cc-edge of the cell of $C_j$ between them.
	\end{itemize}
	In the resulting drawing $\calG$ of $G$ (see \cref{fig:example-assignments}), each \Gg-edge crosses every $\Var_+$-gadget precisely in two \Rr-edges, two \Bb-edges, two \LB-edges and one \Cc-edge.
	By \cref{tab:colors}, we altogether get $\crn(\calG)=\crn(\calG'_+)+n\ell\big(2\omega^6+4\omega^4+\omega^2+\mathcal{O}_{n,\ell}(\omega)\big)$.
	By our choice of $\omega$, we have $\mathcal{O}_{n,\ell}(\omega)\leq(\omega^2-1)$, and hence $\crn(\calG)\leq k$ by \eqref{eq:kvalue}.
\end{proof}

\begin{lemma}\label{lem:redbackward}
	If the constructed instance $(G, k)$ of \textsc{Crossing Number} has a solution $\calG$ with $\crn(\calG)\leq k$,
	then the original \textsc{Satisfiability} instance $\mathcal{I}=(\mathcal{C}, \mathcal{V})$ is satisfiable.
\end{lemma}

\begin{proof}
	We examine the subdrawing $\calG'$ of $G'$ within $\calG$, and assign $\tau$-values of the variables~$x_i$ of~$\mathcal{V}$ for $i\in[n]$ as follows:
	$\tau(x_i)=\true$ if the \texttt{LB-pos} path is to the left of \texttt{LB-neg} in~$\calG'$, and~$\tau(x_i)=\false$ otherwise.
	We claim that $\tau$ is a satisfying truth assignment of the instance~$\mathcal{I}$.
	
	Let $f$ the \Gg-edge of the clause $C_j\in\mathcal{C}$ for $j\in[\ell]$, as drawn in $\calG$.
	By \cref{cor:cellsarea}\,b), there is~$i\in[n]$ such that $f$ crosses the cell of $C_j$ in~$\Var_+^i$ in the ``jump-up'' fashion, that is, crossing~$\Var_+^i$ precisely in the edges $v^i_{4j+i-3,L} v^i_{4j+i-2,L}$ and $v^i_{4j+i-2,R} v^i_{4j+i-1,R}$.
	By \cref{cor:ccedges}, the edge~$f$ can cross only one \Cc-edge within the cell of $C_j$ in $\Var_+^i$.
	Examining the possible drawings of the cell in \cref{fig:cells}, we see that this is possible only if $\tau(x_i)=\true$ and $x_i\in C_j$, or $\tau(x_i)=\false$ and $\overline{x_m}\in C_j$.
	In other words, only if $x_i$ satisfies $C_j$, and this is true for all clauses of the original instance $\mathcal{I}$.
\end{proof}

Since the construction of the instance $(G, k)$ of \textsc{Crossing Number} in \cref{sub:fullred} can clearly be finished in polynomial time, \cref{lem:redforward} and \cref{lem:redbackward} together imply the first part of \cref{thm:mainred} claiming equivalence of the considered instances.

\subsection{On Path-width of the Resulting Instance}\label{subsec:width}

The final ingredient needed to complete the proof of \Cref{thm:mainred} (and thus of \Cref{thm:main}) is an estimate of the path-width and tree-width of the constructed instance~$G$.
To establish this formally, we will leverage the cops-and-robber game characterization of path-width from \Cref{thm:STcops}.

We start with an auxiliary technical claim.
\begin{lemma}\label{lem:sweep}
	Let $H$ be a graph whose vertex set is partitioned into $m$ disjoint parts $V(H)=A_1\cup\ldots\cup A_m$, and for each $i\in\{1,\ldots,m\}$, let $A_i=\{v_{i,j}:a_i\leq j\leq b_i\}$ for some integers $a_i\leq b_i$.  
	Assume that%
	\footnote{Intuitively, the graph $H$ can be easily pictured as having a planar drawing with the paths $A_1, \ldots, A_m$ as vertical columns from left to right,
		and other edges between neighboring columns connecting only vertices on the same level or between two consecutive levels.}
	\begin{enumerate}[a)]
		\item each $A_i$ induces a path in $H$ along the natural vertex order, i.e., $v_{i,a_i},v_{i,a_i+1},\ldots,v_{i,b_i}$;
		\item if an edge $v_{i,j}v_{i',j'}$ exists with $i\not=i'$, then $|i-i'|=1$ and $|j-j'|\leq1$; and
		\item there are no indices $i\not=i'$ and $j\not=j'$ such that both $v_{i,j}v_{i',j'}\in E(H)$ and $v_{i,j'}v_{i',j}\in E(H)$.
	\end{enumerate}
	Then there exists a valid monotone search strategy for the cops on $H$ using $m+1$ cops against an invisible robber.
	Furthermore, this strategy can start with the cops occupying the vertex subset $\{v_{1,a_1},\ldots,v_{m,a_m}\}$.
\end{lemma}

\begin{proof}
	Without loss of generality, we may assume that $a_1=\ldots=a_m=1$ and $b_1=\ldots=b_m$.
	If this is not the case, we can add dummy vertices and edges to $H$ to extend each path $A_i$. So that all cops start at the same index $1$ and end at the same index $b$, and then modify the cop strategy to occupy $v_{i,a_i}$ instead of $v_{i,j}$ for $j<a_i$, and to occupy $v_{i,b_i}$ instead of~$v_{i,j}$ for~$j>b_i$.
	
	We organize the vertices into levels: for each $j\in [b]$, level $j$ consists of the vertices $\{v_{1,j},\ldots,v_{m,j}\}$. 
	As an initial placement, we start the search by placing $m$ cops on $\{v_{1,1},\ldots,v_{m,1}\}$, i.e., the first vertex of each vertical path.
	Let the $(m+1)^{\text{th}}$ cop be used for the sweeping step described below.
	
	We proceed inductively on the level $j$. 
	Assume that at some stage of the search: \emph{(1)} each path $A_i$ contains exactly one cop at level $j$ or $j+1$, forming a set $X\subseteq V(H)$, \emph{(2)} all vertices on levels $\leq j$ are guaranteed robber-free, \emph{(3)} there is no edge in $H-X$ between levels $j$ and~$j+1$, and \emph{(4)} at least one cop still remains on level~$j$.
	
	We now describe the procedure of sweeping to the next level.
	Define a \emph{diagonal-up edge from $v_{i,j}$} as any edge  $v_{i,j}v_{i+1,j}$ or $v_{i,j}v_{i-1,j}$ leading “upward” to level $j+1$.
	Let $i,i'$ be indices such that $i+1<i'$ and the cops occupy $v_{i,j+1}$ and $v_{i',j+1}$ (or $i=0$, or $i'=m+1$ if at boundaries), and all vertices $v_{i+1,j},\ldots,v_{i'-1,j}\in X$.
	
	By the non-crossing condition (c) of the Lemma, there can be at most $i'-i-2$ diagonal-up edges from $v_{i+1,j},\ldots,v_{i'-1,j}$ whose endpoint at level $j+1$ is not already in~$X$.
	Hence, by the pigeonhole principle, there exists $i''$ with $i+1\leq i''\leq i'-1$ such that the only neighbor of~$v_{i''\!,j}$ on level~$j+1$ and not in $X$ is $v_{i''\!,j+1}$.
	We place the remaining $(m+1)^{\text{th}}$ cop on $v_{i''\!,j+1}$ and, subsequently, lift the cop from $v_{i''\!,j}$.
	After this step, the robber cannot move back to~$v_{i''\!,j}$ or any lower level vertices, because all paths are either blocked or already cleared.
	Let the new cop set be $X':=(X\setminus\{v_{i''\!,j}\})\cup\{v_{i''\!,j+1}\}$.
	Clearly, the set $X'$ again satisfies all the inductive the assumptions of the previous paragraph.
	Repeating this process for each level, we sweep all cops upward until level $b$.
	
	Hence, the $m+1$ cops can monotonically clear $H$ starting from $\{v_{1,1},\ldots,v_{m,1}\}$, and eventually capture the robber.
\end{proof}

\begin{proposition}\label{pro:wbound}
	For any given instance of \textsc{Satisfiability}, the graph $G$ constructed in \Cref{sub:fullred} (for the proof of \Cref{thm:mainred}) has path-width at most $12$ and of tree-width at most~$9$.
\end{proposition}

\begin{proof}
	The proof is completed once we find a monotone search strategy for the cop player on $G$ using $13$ cops against an invisible robber (which implies path-width at most $12$), and a strategy using $10$ cops against a visible robber (which implies tree-width at most $9$).
	We start with the path-width bound.
	
	First, place $8$ cops (see \cref{fig:auxill}) on the vertices $$r_{1,L}^1, r_{h+3,L}^1, u^{BL}, u^{TL} \text{(left side of }G\text{) and } r_{1,R}^n, r_{h+3,R}^n, u^{BR}, u^{TR} \text{ (right side of }G\text{)}.$$
	This placement separates the set $U_0\subseteq V(G)$, consisting of the vertices of the left and right \HB-paths of the frame and the \texttt{R-left} path of $\Var_+^1$ and the \texttt{R-right} path of $\Var_+^n$, from the rest of the graph $G$ (see \cref{fig:example-assignments}).
	
	We can now use additional $5$ cops ($m=4$ plus one extra) to search the subgraph induced by~$U_0$ using \cref{lem:sweep}.
	Notice, in this application, the levels in \cref{lem:sweep} are shifted relative to the natural vertex indexing from the construction of~$G$.
	
	After this initial phase, the search continues inductively for $i=1,2,\ldots,n$.
	We assume that $8$ cops occupy $$r_{1,L}^i, r_{h+3,L}^i, u_0^{i-1}, w_0^{i-1} \text{(the latter two being }u^{BL}, u^{TL} \text{ if (}i=1\text{) and also }r_{1,R}^n, r_{h+3,R}^n, u^{BR}, u^{TR}.$$
	Further, we assume that $V(\Var_+^{i-1})$ (if $i>1$) has already been cleared of the robber.
	
	Next, we place $4$ of the remaining cops on vertices $u_0^i$, $u_1^i$, $w_1^i$, $w_0^i$, and subsequently lift the cops from $r_{1,L}^i$, $r_{h+3,L}^i$, $u_0^{i-1}$, $w_0^{i-1}$.
	The $5$ free cops can then search the \texttt{B-/LB-pos} and \texttt{B-/LB-neg} paths of $\Var_+^i$ using \cref{lem:sweep}.
	
	We place $2$ of the free cops on $r_{1,L}^{i+1}$, $r_{h+3,L}^{i+1}$ and use the remaining $3$ cops to search the \texttt{R-right} path of~$\Var_+^i$ and the \texttt{R-left} path of $\Var_+^{i+1}$, again via \cref{lem:sweep}.
	After finishing previous, we lift the cops from~$u_1^i$ and~$w_1^i$ and we are back to the induction assumption with~$i+1$ instead of~$i$.
	
	The described procedure is a valid monotone search strategy against an invisible robber and establishes that~$G$ has path-width at most~$12$.
	\medskip
	
	Regarding the tree-width bound, we slightly improve the strategy by using the robber's visibility.
	After placing the initial 8 cops as above on $$r_{1,L}^1, r_{h+3,L}^1, u^{BL}, u^{TL} \text{ and } r_{1,R}^n, r_{h+3,R}^n, u^{BR}, u^{TR},$$
	we check whether the robber is trapped inside the set $U_0$.
	If this is the case, we place the $9^{\text{th}}$ cop to capture the robber in $U_0$ using \cref{lem:sweep}, together with the $4$ initially on $r_{1,L}^1$, $u^{BL}$, $r_{1,R}^n$, $u^{BR}$.
	
	In the induction phase, whenever the robber is trapped inside the \texttt{B-/LB-pos} and \texttt{B-/LB-neg} paths of~$\Var_+^i$, we can use the $4$ cops from $r_{1,R}^n$, $r_{h+3,R}^n$, $u^{BR}$, $u^{TR}$ to capture the robber via \cref{lem:sweep}.
	The same applies to the \texttt{R-right} path of $\Var_+^i$ and the \texttt{R-left} path of $\Var_+^{i+1}$.
	
	When moving cops from $r_{1,L}^i$, $r_{h+3,L}^i$, $u_0^{i-1}$, $w_0^{i-1}$ to next $u_0^i$, $u_1^i$, $w_1^i$, $w_0^i$, we move in pairs: first place cops on $u_0^i$, $u_1^i$ and lift from $r_{1,L}^i$, $u_0^{i-1}$, and then handle the other pair.
	The maximum number of cops required in this strategy is $4+4+2=10$, which occurs just before trapping the robber in the \texttt{R-right}/\texttt{R-left} paths.
	This shows that the tree-width of $G$ is at most $9$.
\end{proof}

With \cref{pro:wbound}, we have finished the whole proof of \cref{thm:mainred}.

\section{Conclusion}
\label{sec:conclu}
We have shown that the \textsc{Crossing Number} problem is \NP-hard for graphs of path-width~$12$, and consequently also for graphs of tree-width~$9$, thereby resolving a long-standing open problem in the crossing number research.
It is worth noting that parameters such as the clique-width and the rank-width, being bounded by $\mathcal{O}(2^{\tw})$, also yield width decompositions that are too general to be directly useful for tackling the \textsc{Crossing Number} problem.

Possible future research may focus on determining the largest value $t$ such that the \textsc{Crossing Number} problem can be solved in polynomial time on graphs of path-width (or tree-width) at most~$t$.
There, polynomial-time solvability is trivial for $t=2$ (the graphs are planar), and partial results exist for the path-width~$t=3$~\cite{DBLP:journals/algorithmica/BiedlCDM20}.

On the other hand, more restrictive structural parameterizations---such as the treedepth, distance to a linear forest (i.e., distance to union of paths), the feedback vertex set number (distance to a forest), cut-width, or bandwidth---may offer greater potential.
Very few existing crossing-number results extend to these parameters, making it a promising direction to explore whether any of them could lead to either fixed-parameter tractability or \W-hardness results for \textsc{Crossing Number}.
However, nontriviality of computing the exact crossing number on graphs of bounded vertex cover number \cite{DBLP:conf/gd/HlinenyS19} warns that this could be a tough goal.

Note that, since the \textsc{Crossing Number} is known to be fixed-parameter tractable when parameterized by the solution value $k$~\cite{DBLP:journals/jcss/Grohe04,cm-faeg2-21,DBLP:conf/soda/LokshtanovP0S0Z25}, it is meaningful to investigate only parameters that do \emph{not} already bound the crossing number itself.

\bibliography{references}

\begin{thebibliography}{10}

\bibitem{DBLP:journals/algorithmica/BiedlCDM20}
Therese Biedl, Markus Chimani, Martin Derka, and Petra Mutzel.
\newblock Crossing number for graphs with bounded pathwidth.
\newblock {\em Algorithmica}, 82(2):355--384, 2020.
\newblock \href {https://doi.org/10.1007/S00453-019-00653-X}
  {\path{doi:10.1007/S00453-019-00653-X}}.

\bibitem{Cabello13}
Sergio Cabello.
\newblock Hardness of approximation for crossing number.
\newblock {\em Discrete Comput. Geom.}, 49(2):348--358, March 2013.
\newblock \href {https://doi.org/10.1007/S00454-012-9440-6}
  {\path{doi:10.1007/S00454-012-9440-6}}.

\bibitem{DBLP:journals/algorithmica/CabelloM11}
Sergio Cabello and Bojan Mohar.
\newblock Crossing number and weighted crossing number of near-planar graphs.
\newblock {\em Algorithmica}, 60(3):484--504, 2011.
\newblock \href {https://doi.org/10.1007/S00453-009-9357-5}
  {\path{doi:10.1007/S00453-009-9357-5}}.

\bibitem{DBLP:journals/siamcomp/CabelloM13}
Sergio Cabello and Bojan Mohar.
\newblock Adding one edge to planar graphs makes crossing number and
  1-planarity hard.
\newblock {\em {SIAM} J. Comput.}, 42(5):1803--1829, 2013.
\newblock \href {https://doi.org/10.1137/120872310}
  {\path{doi:10.1137/120872310}}.

\bibitem{DBLP:journals/jco/ChimaniH17}
Markus Chimani and Petr Hlin\v{e}n{\'{y}}.
\newblock A tighter insertion-based approximation of the crossing number.
\newblock {\em J. Comb. Optim.}, 33(4):1183--1225, 2017.
\newblock \href {https://doi.org/10.1007/S10878-016-0030-Z}
  {\path{doi:10.1007/S10878-016-0030-Z}}.

\bibitem{DBLP:journals/jct/ChimaniHS20}
Markus Chimani, Petr Hlin\v{e}n{\'{y}}, and Gelasio Salazar.
\newblock Toroidal grid minors and stretch in embedded graphs.
\newblock {\em J. Comb. Theory, Ser. {B}}, 140:323--371, 2020.
\newblock \href {https://doi.org/10.1016/J.JCTB.2019.05.009}
  {\path{doi:10.1016/J.JCTB.2019.05.009}}.

\bibitem{DBLP:conf/focs/ChuzhoyMT20}
Julia Chuzhoy, Sepideh Mahabadi, and Zihan Tan.
\newblock Towards better approximation of graph crossing number.
\newblock In {\em 61st {IEEE} Annual Symposium on Foundations of Computer
  Science, {FOCS} 2020}, pages 73--84. {IEEE}, 2020.
\newblock \href {https://doi.org/10.1109/FOCS46700.2020.00016}
  {\path{doi:10.1109/FOCS46700.2020.00016}}.

\bibitem{DBLP:conf/stoc/ChuzhoyT22}
Julia Chuzhoy and Zihan Tan.
\newblock A subpolynomial approximation algorithm for graph crossing number in
  low-degree graphs.
\newblock In {\em {STOC} '22: 54th Annual {ACM} {SIGACT} Symposium on Theory of
  Computing}, pages 303--316. {ACM}, 2022.
\newblock \href {https://doi.org/10.1145/3519935.3519984}
  {\path{doi:10.1145/3519935.3519984}}.

\bibitem{cm-faeg2-21}
{\'{E}}ric {Colin de Verdi{\`{e}}re}, {\'E}ric, and Thomas Magnard.
\newblock An {FPT} algorithm for the embeddability of graphs into
  two-dimensional simplicial complexes.
\newblock In {\em Proceedings of the 29th European Symposium on Algorithms
  (ESA)}, pages 32:1--32:17, 2021.
\newblock \href {https://doi.org/10.4230/LIPICS.ESA.2021.32}
  {\path{doi:10.4230/LIPICS.ESA.2021.32}}.

\bibitem{DBLP:conf/esa/VerdiereH25}
{\'{E}}ric {Colin de Verdi{\`{e}}re} and Petr Hlin\v{e}n{\'{y}}.
\newblock A unified {FPT} framework for crossing number problems.
\newblock In {\em 33rd Annual European Symposium on Algorithms, {ESA} 2025},
  volume 351 of {\em LIPIcs}, pages 21:1--21:18. Schloss Dagstuhl -
  Leibniz-Zentrum f{\"{u}}r Informatik, 2025.
\newblock \href {https://doi.org/10.4230/LIPICS.ESA.2025.21}
  {\path{doi:10.4230/LIPICS.ESA.2025.21}}.

\bibitem{CyganFKLMPPS15}
Marek Cygan, Fedor~V. Fomin, Lukasz Kowalik, Daniel Lokshtanov, D{\'{a}}niel
  Marx, Marcin Pilipczuk, Michal Pilipczuk, and Saket Saurabh.
\newblock {\em Parameterized Algorithms}.
\newblock Springer, 2015.
\newblock \href {https://doi.org/10.1007/978-3-319-21275-3}
  {\path{doi:10.1007/978-3-319-21275-3}}.

\bibitem{BattistaETT99}
Giuseppe {Di Battista}, Peter Eades, Roberto Tamassia, and Ioannis~G. Tollis.
\newblock {\em Graph Drawing: Algorithms for the Visualization of Graphs}.
\newblock Prentice-Hall, 1999.

\bibitem{Diestel12}
Reinhard Diestel.
\newblock {\em Graph Theory, 4th Edition}, volume 173 of {\em Graduate texts in
  mathematics}.
\newblock Springer, 2012.
\newblock \href {https://doi.org/10.1007/978-3-662-53622-3}
  {\path{doi:10.1007/978-3-662-53622-3}}.

\bibitem{GareyJ83}
Michael~R. Garey and David~S. Johnson.
\newblock Crossing number is {NP-complete}.
\newblock {\em SIAM J. Algebr. Discrete Methods}, 4(3):312--316, September
  1983.
\newblock \href {https://doi.org/10.1137/0604033} {\path{doi:10.1137/0604033}}.

\bibitem{DBLP:journals/combinatorics/GitlerHLS08}
Isidoro Gitler, Petr Hlin\v{e}n{\'{y}}, Jes{\'{u}}s Lea{\~{n}}os, and Gelasio
  Salazar.
\newblock The crossing number of a projective graph is quadratic in the
  face-width.
\newblock {\em Electron. J. Comb.}, 15(1), 2008.
\newblock \href {https://doi.org/10.37236/770} {\path{doi:10.37236/770}}.

\bibitem{DBLP:journals/jcss/Grohe04}
Martin Grohe.
\newblock Computing crossing numbers in quadratic time.
\newblock {\em J. Comput. Syst. Sci.}, 68(2):285--302, 2004.
\newblock \href {https://doi.org/10.1016/J.JCSS.2003.07.008}
  {\path{doi:10.1016/J.JCSS.2003.07.008}}.

\bibitem{DBLP:conf/compgeom/HammH22}
Thekla Hamm and Petr Hlin\v{e}n{\'{y}}.
\newblock Parameterised partially-predrawn crossing number.
\newblock In {\em 38th International Symposium on Computational Geometry, SoCG
  2022}, volume 224 of {\em LIPIcs}, pages 46:1--46:15. Schloss Dagstuhl -
  Leibniz-Zentrum f{\"{u}}r Informatik, 2022.
\newblock \href {https://doi.org/10.4230/LIPICS.SOCG.2022.46}
  {\path{doi:10.4230/LIPICS.SOCG.2022.46}}.

\bibitem{DBLP:journals/jct/Hlineny06a}
Petr Hlin\v{e}n{\'{y}}.
\newblock Crossing number is hard for cubic graphs.
\newblock {\em J. Comb. Theory, Ser. {B}}, 96(4):455--471, 2006.
\newblock \href {https://doi.org/10.1016/j.jctb.2005.09.009}
  {\path{doi:10.1016/j.jctb.2005.09.009}}.

\bibitem{DBLP:conf/mfcs/Hlineny25}
Petr Hlin\v{e}n{\'{y}}.
\newblock Complexity of anchored crossing number and crossing number of almost
  planar graphs.
\newblock In {\em 50th International Symposium on Mathematical Foundations of
  Computer Science, {MFCS} 2025}, volume 345 of {\em LIPIcs}, pages
  59:1--59:17. Schloss Dagstuhl - Leibniz-Zentrum f{\"{u}}r Informatik, 2025.
\newblock \href {https://doi.org/10.4230/LIPICS.MFCS.2025.59}
  {\path{doi:10.4230/LIPICS.MFCS.2025.59}}.

\bibitem{DBLP:conf/compgeom/HlinenyD16}
Petr Hlin\v{e}n{\'{y}} and Marek Der\v{n}{\'{a}}r.
\newblock Crossing number is hard for kernelization.
\newblock In {\em 32nd International Symposium on Computational Geometry, SoCG
  2016}, volume~51 of {\em LIPIcs}, pages 42:1--42:10. Schloss Dagstuhl -
  Leibniz-Zentrum f{\"{u}}r Informatik, 2016.
\newblock \href {https://doi.org/10.4230/LIPICS.SOCG.2016.42}
  {\path{doi:10.4230/LIPICS.SOCG.2016.42}}.

\bibitem{HK-isaac24}
Petr Hlin\v{e}n\'{y} and Liana Khazaliya.
\newblock {Crossing Number Is NP-Hard for Constant Path-Width (And
  Tree-Width)}.
\newblock In {\em 35th International Symposium on Algorithms and Computation
  (ISAAC 2024)}, volume 322 of {\em Leibniz International Proceedings in
  Informatics (LIPIcs)}, pages 40:1--40:15, Dagstuhl, Germany, 2024. Schloss
  Dagstuhl -- Leibniz-Zentrum f{\"u}r Informatik.
\newblock \href {https://doi.org/10.4230/LIPIcs.ISAAC.2024.40}
  {\path{doi:10.4230/LIPIcs.ISAAC.2024.40}}.

\bibitem{DBLP:conf/gd/HlinenyS06}
Petr Hlin\v{e}n{\'{y}} and Gelasio Salazar.
\newblock On the crossing number of almost planar graphs.
\newblock In {\em Graph Drawing, 14th International Symposium, {GD} 2006},
  volume 4372 of {\em Lecture Notes in Computer Science}, pages 162--173.
  Springer, 2006.
\newblock \href {https://doi.org/10.1007/978-3-540-70904-6\_17}
  {\path{doi:10.1007/978-3-540-70904-6\_17}}.

\bibitem{DBLP:conf/isaac/HlinenyS15}
Petr Hlin\v{e}n{\'{y}} and Gelasio Salazar.
\newblock On hardness of the joint crossing number.
\newblock In {\em {ISAAC}}, volume 9472 of {\em Lecture Notes in Computer
  Science}, pages 603--613. Springer, 2015.
\newblock \href {https://doi.org/10.1007/978-3-662-48971-0_51}
  {\path{doi:10.1007/978-3-662-48971-0_51}}.

\bibitem{DBLP:conf/gd/HlinenyS19}
Petr Hlin\v{e}n{\'{y}} and Abhisekh Sankaran.
\newblock Exact crossing number parameterized by vertex cover.
\newblock In {\em Graph Drawing and Network Visualization - 27th International
  Symposium, {GD} 2019}, volume 11904 of {\em Lecture Notes in Computer
  Science}, pages 307--319. Springer, 2019.
\newblock \href {https://doi.org/10.1007/978-3-030-35802-0\_24}
  {\path{doi:10.1007/978-3-030-35802-0\_24}}.

\bibitem{DBLP:conf/stoc/KawarabayashiR07}
Ken{-}ichi Kawarabayashi and Bruce~A. Reed.
\newblock Computing crossing number in linear time.
\newblock In {\em Proceedings of the 39th Annual {ACM} Symposium on Theory of
  Computing, STOC 2007}, pages 382--390. {ACM}, 2007.
\newblock \href {https://doi.org/10.1145/1250790.1250848}
  {\path{doi:10.1145/1250790.1250848}}.

\bibitem{DBLP:conf/soda/LokshtanovP0S0Z25}
Daniel Lokshtanov, Fahad Panolan, Saket Saurabh, Roohani Sharma, Jie Xue, and
  Meirav Zehavi.
\newblock Crossing number in slightly superexponential time (extended
  abstract).
\newblock In {\em Proceedings of the 2025 Annual {ACM-SIAM} Symposium on
  Discrete Algorithms, {SODA} 2025}, pages 1412--1424. {SIAM}, 2025.
\newblock \href {https://doi.org/10.1137/1.9781611978322.44}
  {\path{doi:10.1137/1.9781611978322.44}}.

\bibitem{DBLP:conf/gd/PelsmajerSS07a}
Michael~J. Pelsmajer, Marcus Schaefer, and Daniel Stefankovic.
\newblock Crossing numbers and parameterized complexity.
\newblock In {\em Graph Drawing, 15th International Symposium, {GD} 2007},
  volume 4875 of {\em Lecture Notes in Computer Science}, pages 31--36.
  Springer, 2007.
\newblock \href {https://doi.org/10.1007/978-3-540-77537-9\_6}
  {\path{doi:10.1007/978-3-540-77537-9\_6}}.

\bibitem{riskin1996crossing}
Adrian Riskin.
\newblock The crossing number of a cubic plane polyhedral map plus an edge.
\newblock {\em Studia Scientiarum Mathematicarum Hungarica}, 31(4):405--414,
  1996.

\bibitem{DBLP:journals/jct/SeymourT93}
Paul~D. Seymour and Robin Thomas.
\newblock Graph searching and a min-max theorem for tree-width.
\newblock {\em J. Comb. Theory {B}}, 58(1):22--33, 1993.
\newblock \href {https://doi.org/10.1006/jctb.1993.1027}
  {\path{doi:10.1006/jctb.1993.1027}}.

\bibitem{Turan77}
Paul Turán.
\newblock A note of welcome.
\newblock {\em Journal of Graph Theory}, 1(1):7--9, 1977.
\newblock \href {https://doi.org/10.1002/jgt.3190010105}
  {\path{doi:10.1002/jgt.3190010105}}.

\bibitem{DBLP:journals/csr/Zehavi22}
Meirav Zehavi.
\newblock Parameterized analysis and crossing minimization problems.
\newblock {\em Comput. Sci. Rev.}, 45:100490, 2022.
\newblock \href {https://doi.org/10.1016/J.COSREV.2022.100490}
  {\path{doi:10.1016/J.COSREV.2022.100490}}.

\end{thebibliography}


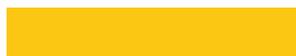
\end{document}